\documentclass[final,10pt,twocolumn,superscriptaddress]{revtex4-2}
\usepackage{amsmath,amssymb,bm}
\usepackage{graphicx}
\usepackage{amstext}
\usepackage{amsthm}
\usepackage{latexsym}
\usepackage{microtype}
 
\usepackage[T1]{fontenc}
\usepackage{dcolumn}
\usepackage{bm}
\usepackage[colorlinks=true,allcolors=blue]{hyperref}
\usepackage{physics}
\usepackage{siunitx}
\usepackage[version=4]{mhchem}
\usepackage{bbm}
\usepackage[usenames,dvipsnames]{xcolor}

\makeatletter
\def\frutiger{cmss10 }
\def\frutigerbold{cmssbx10 }
=\frutigerbold at 14pt
=\frutiger at 8pt
=\frutigerbold at 12pt
=\frutigerbold at10pt
=\frutigerbold at 8pt
=\frutiger at 8pt
=\frutigerbold at 31pt
\def\@caption@tabnum@sep{\figtextfont{{ }{\bf\textbar}{ }}}%
\def\fnum@table{{\bf\tablename~\thetable}}

\def\@caption@fignum@sep{\figtextfont{{ }{\bf\textbar}{ }}}%
\renewenvironment{figure}{\@float{figure}\def\textbf##1{{\fignumfont ##1}}\def\bf{\fignumfont}}{\end@float}
\renewcommand{\fnum@figure}{\bf Fig. \thefigure}
\def\@startsection#1#2#3#4#5#6{%
\if@noskipsec\leavevmode\fi
\par\@tempskipa #4\relax
\@afterindenttrue
\ifdim\@tempskipa <\z@
\@tempskipa -\@tempskipa \@afterindentfalse
\fi\if@nobreak\everypar{}%
\else\addpenalty\@secpenalty\addvspace\@tempskipa\fi
\@ifstar{\@ssect{#3}{#4}{#5}{#6}}{\@dblarg{\@sect{#1}{#2}{#3}{#4}{#5}{#6}}}}
\def\@sect#1#2#3#4#5#6[#7]#8{%
\ifnum #2>0
\let\@svsec\@empty
\else\refstepcounter{#1}\protected@edef\@svsec{\@seccntformat{#1}\relax}\fi
\@tempskipa #5\relax
\ifdim\@tempskipa>\z@
\begingroup#6{\@hangfrom{\hskip #3\relax\@svsec}%
\interlinepenalty \@M #8\@@par}\endgroup
\csname #1mark\endcsname{#7}%
\addcontentsline{toc}{#1}{%
\ifnum #2>\c@secnumdepth\else
\protect\numberline{\csname the#1\endcsname}\fi #7}%
\else\def\@svsechd{#6{\hskip #3\relax
\@svsec #8\ifnum#2=2.\fi}%
\csname #1mark\endcsname{#7}%
\addcontentsline{toc}{#1}{%
\ifnum #2>\c@secnumdepth \else
\protect\numberline{\csname the#1\endcsname}\fi #7}}%
\fi\@xsect{#5}}
\renewcommand\section{\@startsection {section}{1}{\z@}%
{-10pt \@plus -1ex \@minus -.2ex}{.5ex }{\normalfont\Large\bfseries\sectionfont}}
\renewcommand\subsection{\@startsection{subsection}{2}{\z@}%
{10pt\@plus 1ex \@minus .2ex}{-0.5ex \@plus .2ex}{\normalfont\large\bfseries\subsectionfont}}
\def\frontmatter@title@format{\titlefont\centering}%
\def\frontmatter@title@below{\addvspace{-5pt}}%

\renewenvironment{thebibliography}[1]{%
\bib@heading%
  \ifx\bibpreamble\relax\else\ifx\bibpreamble\@empty\else
    \noindent\bibpreamble\par\nobreak
  \fi\fi
  \list{\@biblabel{\@arabic\c@enumiv}}%
  {\settowidth\labelwidth{\@biblabel{#1}}%
    \leftmargin\labelwidth
    \advance\leftmargin\labelsep
    \@openbib@code
    \usecounter{enumiv}%
    \let\p@enumiv\@empty
    \renewcommand*\theenumiv{\@arabic\c@enumiv}
}%
  \sloppy\clubpenalty4000\widowpenalty4000%
  \sfcode`\.=\@m}
{\def\@noitemerr
  {\@latex@warning{Empty `thebibliography' environment}}%
  \endlist}
\newcommand*\bib@heading{%
  \section{\refname}
  \fontsize{8}{10}\selectfont
}
\newcommand*\@openbib@code{%
      \advance\leftmargin\bibindent
      \itemindent -\bibindent
      \listparindent \itemindent
      \parsep \z@
}%
\newdimen\bibindent
\makeatother

\begin{document}

\preprint{APS/123-QED}

\title{Efficient calculation of carrier scattering rates from first principles}

\author{Alex M. Ganose}
\email{aganose@lbl.gov}
\affiliation{Energy Technologies Area, Lawrence Berkeley National Laboratory, Berkeley, California 94720, USA}

\author{Junsoo Park}
\affiliation{Energy Technologies Area, Lawrence Berkeley National Laboratory, Berkeley, California 94720, USA}

\author{Alireza Faghaninia}
\affiliation{Energy Technologies Area, Lawrence Berkeley National Laboratory, Berkeley, California 94720, USA}

\author{Rachel Woods-Robinson}
\affiliation{Energy Technologies Area, Lawrence Berkeley National Laboratory, Berkeley, California 94720, USA}
\affiliation{Department of Materials Science and Engineering, University of California Berkeley, California 94720, United States}

\author{Kristin A. Persson}
\affiliation{Department of Materials Science and Engineering, University of California Berkeley, California 94720, United States}
\affiliation{Molecular Foundry, Energy Sciences Area,  Lawrence Berkeley National Laboratory, Berkeley, California 94720, USA}

\author{Anubhav Jain}
\email{ajain@lbl.gov}
\affiliation{Energy Technologies Area, Lawrence Berkeley National Laboratory, Berkeley, California 94720, USA}

\date{\today}

\begin{abstract}
The electronic transport behaviour of materials determines their suitability for technological applications.
We develop a computationally efficient method for calculating carrier scattering rates of solid-state semiconductors
and insulators from first principles inputs. The present method extends existing polar and non-polar electron-phonon
coupling, ionized impurity, and piezoelectric scattering mechanisms formulated for
isotropic band structures to support highly anisotropic materials. We test the formalism by 
calculating the electronic transport properties of 19 semiconductors, including the large 48 atom CH$_3$NH$_3$PbI$_3$ hybrid perovskite, 
and comparing the results against experimental measurements and more detailed scattering simulations.
The Spearman rank coefficient of mobility against experiment ($r_\mathrm{s} = $ 0.92) improves significantly on 
results obtained using a constant relaxation time approximation ($r_\mathrm{s} = $ 0.50).
We find our approach offers similar accuracy to state-of-the art methods at approximately 1/500th the computational cost, thus enabling its use
in high-throughput computational workflows for the accurate screening of carrier mobilities, lifetimes, and thermoelectric power. 
\end{abstract}

\maketitle

\section{Introduction}

Solid-state materials exhibit a variety of electronic transport behaviors, enabling their deployment in a variety of technological applications, including light-emitting devices, photocatalysts, transparent conductors, solar cells, and thermoelectrics \cite{pimputkar2009prospects,green2014emergence,fujishima1972electrochemical,jain2016computational,ellmer2012past,snyder2008complex}.
Recent years have seen an explosion of interest into the computational prediction of electronic transport properties, leading to a hierarchy of methods that that can be broadly split into three categories.
(i) Semi-empirical models for approximating electron lifetimes have been employed since the 1930s \cite{bloch1929quantenmechanik,bardeen1950deformation,herring1956transport,harrison1956scattering,meijer1953note,frohlich1954electrons} but have seen a resurgence with the advent of large-scale materials science databases due to their computational efficiency \cite{xi2018discovery,yan2015MaterialDescriptors,wang2011AssessingThermoelectric}.
These approaches have recently been extended to permit first-principles inputs \cite{faghaninia2015amobt,m2019ammcr,chaves2020investigating,li2021TransOptCode} but the underlying assumption of single parabolic bands with no anisotropy limits their widespread application \cite{long1959IonizedImpurityScattering}.
(ii) The second category eschews the calculation of electron lifetimes, instead employing a constant scattering rate for all electronic states.
When combined with Fourier \cite{madsen2006boltztrap,madsen2018boltztrap2} or Wannier \cite{pizzi2014boltzwann} interpolation of \textit{ab initio} electronic band structures this enables efficient calculation of transport properties in complex systems with multiple non-parabolic bands \cite{isaacs2018InverseBand, madsen2006AutomatedSearch,bhattacharya2016NovelPtype}.
Recent work has applied this approach to compute the transport behaviour of large numbers of materials, including 48,000 semiconductors in the Materials Project database by \citet{ricci2017ab}, 809 sulfides by \citet{miyata2018HighThroughputScreening}, and 75 potential thermoelectric candidates by \citet{xing2017ElectronicFitness}; however, the unphysical treatment of electron scattering and the reliance on a empirical tuning parameter often results in significant errors.
(iii) Finally, fully-first principles approaches to calculating the electron-phonon interaction based on density functional perturbation theory (DFPT) combined with Wannier interpolation can now yield highly accurate electron lifetimes and have demonstrated remarkable agreement to experimental measurements of electron mobility and conductivity \cite{ponce2016epw,agapito2018ab,ponce2018towards,gonze2020abinit,brunin2020phonon,zhou2020perturbo}.
The calculation of the scattering matrix elements needed to obtain electron lifetimes is extremely computationally demanding, even when approximations are made.
With few exceptions \cite{ghosh2016InitioCalculation,ponce2019OriginLow,li2019DimensionalCrossover}, such approaches have been applied to highly symmetric systems with limited numbers of atoms. \cite{samsonidze2018accelerated,ponce2020review,ma2018IntrinsicPhononlimited,ponce2019HoleMobility,cao2018DominantElectronphonon,zhou2016InitioElectron}.
Although the computational cost of mobility calculations can be reduced though energy-averaging of the matrix elements \cite{deng2020EPICSTAR}, the initial DFPT calculation needed to obtain the matrix elements typically represents the majority of the computational expense.
Despite the range of computational techniques available, no existing method can be applied to compute the transport properties of a broad array of complex materials both accurately and inexpensively [Fig.~(\ref{fig:timing})].
This limitation is a primary obstacle in the application of high-throughput computations to the search for novel functional materials as well as applying this theory to larger and more complex materials.

In the present work, we develop an efficient formalism for calculating anisotropic transport properties of semiconductors that is accurate over a range of materials and amenable to use in high-throughput computational workflows.
Our approach relies on inputs that can be obtained from low-cost \textit{ab initio} methods and that are routinely available in computational materials science databases.
Scattering rates are calculated using the momentum relaxation time approximation (MRTA) to the Boltzmann transport equation (BTE).
The present method includes fully anisotropic acoustic deformation potential, piezoelectric, ionized impurity, and polar electron-phonon scattering.
As an initial test of the approach, we calculate the temperature-dependent electron mobility and Seebeck coefficient of 19 semiconductors including the large 48-atom CH$_3$NH$_3$PbI$_3$ hybrid perovskite.
The Spearman rank coefficient of mobility against experiment ($r_\mathrm{s} = $ 0.92) improves significantly on 
results obtained using a constant relaxation time approximation ($r_\mathrm{s} = $ 0.50).
Furthermore, we find our approach offers similar accuracy to state-of-the art methods at 1/500th the computational cost.
An open source software implementation of the method is made freely available.

\section{Results}

\subsection{Computationally efficient matrix elements}

In the Boltzmann transport equation, the scattering rate of an electron from an initial state $n\mathbf{k}$, where $n$ is a band index and $\mathbf{k}$ is a wave vector, to final state $m\mathbf{k}+\mathbf{q}$ is described by Fermi's golden rule as
\begin{equation}
\begin{aligned}
    \tau_{n\mathbf{k}\rightarrow m\mathbf{k}+\mathbf{q}}^{-1} =
        \frac{2\pi}{\hbar} &{}  \lvert g_{nm}(\mathbf{k}, \mathbf{q}) \rvert^2 \delta \left ( \varepsilon_{n\mathbf{k}} -  \varepsilon_{m\mathbf{k} + \mathbf{q}} \right ),
\end{aligned}
\label{eqn:rate}
\end{equation}
where $\hbar$ is the reduced Planck constant, $\varepsilon$ is the electron energy, $\delta$ is the Dirac delta function and $g$ is the coupling matrix element.
The above equation is given for the case of perfectly elastic scattering \cite{rode1975low}, in which no energy is gained or lost during the scattering process.
A similar equation can be defined for inelastic processes (see Sec.~I of the Supplemental Material), for instance to describe scattering that occurs via emission or absorption of a phonon.
In the constant relaxation time approximation (CRTA), Eq.~(\ref{eqn:rate}) is simplified to a single constant.
In general, however, the impact of different  scattering mechanisms is expressed via the coupling matrix element $g_{nm}(\mathbf{k}, \mathbf{q}) = \mel{m\mathbf{k}+\mathbf{q}}{\Delta_\mathbf{q}V}{n\mathbf{k}}$ where $\Delta_\mathbf{q}V$ is an electronic perturbation of some kind.
The primary obstacle in obtaining accurate transport properties is evaluating $g_{nm}(\mathbf{k}, \mathbf{q})$ on extremely dense Brillouin zone grids, which has so far proven computationally prohibitive for all but the simplest systems \cite{giustino2007electron,giustino2017electron}.

Historically, this challenge has been avoided by use of model matrix elements formulated for isotropic band structures using intrinsic materials parameters.
For example, the treatment of deformation potential scattering due to long-wavelength acoustic phonons proposed by \citet{bardeen1950deformation} depends only on an averaged elastic constant and band edge deformation potential; it ignores perturbations from transverse phonon modes and anisotropy in the deformation response. This simple approach has been employed widely in computations of acoustic phonon scattering but is unreliable and does not generalise to complex systems or metals \cite{khan1984DeformationPotentials,kartheuser1986DeformationPotentials,resta1991DeformationpotentialTheorem}.
An alternative approach, developed by \citet{khan1984DeformationPotentials}, can reproduce the fully-first principles electron-phonon scattering rate if the strain tensor caused by the phonon and an additional velocity term are included.
The resulting matrix element is given by
\begin{equation}
g_{nm}^\mathrm{KA}
= \mel{m\mathbf{k}+\mathbf{q}}{\mathbf{S}_\mathbf{q}\mathbin{:}(\mathbf{D}_{n\mathbf{k}} + \mathbf{v}_{n\mathbf{k}} \otimes \mathbf{v}_{n\mathbf{k}})}{{n\mathbf{k}}},
\end{equation}
where $:$ denotes the double dot product, $\mathbf{S}_\mathbf{q}$ is the strain associated with an acoustic phonon, $\mathbf{D}_{n\mathbf{k}}$ is the second rank deformation potential tensor and $\mathbf{v}_{n\mathbf{k}}$ is the group velocity.
The velocity term is essential to correct the deformation potential in metals and at states away from the valence or conduction band edge in semiconductors.
In practice, however, this equation is no longer simple to evaluate as it requires knowledge of the atomic displacements (the polarization direction) of the phonon mode in order to obtain the strain tensor.

In the present work, we combine the simplicity of the Bardeen and Shockley approach with the accuracy of the Khan and Allen matrix element by exploiting the acoustoelastic properties of materials.
The dispersion relations for acoustic waves are contained in the Christoffel equation \cite{auld1973acoustic}
\begin{equation}
\left [ \Gamma_\mathbf{\hat{q}} - \rho c^2 \mathbbm{1} \right]\mathbf{\hat{u}} = 0,
\end{equation}
where $\mathbbm{1}$ is the identity matrix, $\mathbf{\hat{q}}$ and $\mathbf{\hat{u}}$ are unit vectors giving the direction of phonon propagation and polarization, respectively, $\rho$ is the density, $c$ is the wave velocity, and $\Gamma_\mathbf{\hat{q}} = \mathbf{C} \mathbf{\hat{q}} \cdot \mathbf{\hat{q}}$ is the Christoffel matrix where $\mathbf{C}$ is the rank 4 elastic constant tensor.
Solving the Christoffel equation for a phonon wave vector direction ($\mathbf{\hat{q}}$) results in three sets of eigenvalues ($\rho c^2$) and eigenvectors ($\mathbf{\hat{u}}$), that correspond to the (quasi-)longitudinal and (quasi-)transverse normal modes of the material.
The unit strain associated with each mode is given by $\mathbf{\hat{S}} = \mathbf{\hat{q}}\otimes\mathbf{\hat{u}}$ and the amplitude of the strain at any temperature $T$ can be obtained from the potential energy of the acoustic phonon as $\sqrt{k_\mathrm{B} T/ \rho c^2}$, where $k_\mathrm{B}$ is the Boltzmann constant \cite{zook1964PiezoelectricScattering}.
From this we arrive at an expression for acoustic deformation potential scattering (``ad'') that relies only on the deformation potentials and elastic constants and includes scattering from longitudinal and transverse modes in a single matrix element, given in the Born approximation \cite{born1926quantenmechanik} as
\begin{widetext}
\begin{equation}
g_{nm}^\mathrm{ad}(\mathbf{k}, \mathbf{q}) = 
   \sqrt{k_\mathrm{B} T}  \sum_{\mathbf{G} \neq -\mathbf{q}} \left[ 
        \frac{\mathbf{\tilde{D}}_{n\mathbf{k}} \mathbin{:} \mathbf{\hat{S}}_l}{c_l\sqrt{\rho}} + 
        \frac{\mathbf{\tilde{D}}_{n\mathbf{k}} \mathbin{:} \mathbf{\hat{S}}_{t_1}}{c_{t_1}\sqrt{\rho}} + 
        \frac{\mathbf{\tilde{D}}_{n\mathbf{k}} \mathbin{:} \mathbf{\hat{S}}_{t_2}}{c_{t_2}\sqrt{\rho}}
    \right] \mel{m\mathbf{k}+\mathbf{q}}{e^{i(\mathbf{q} + \mathbf{G})\cdot\mathbf{r}}}{n\mathbf{k}}
\label{eqn:element_ad}
\end{equation}
\end{widetext}
where $\mathbf{\tilde{D}}_{n\mathbf{k}} = \mathbf{D}_{n\mathbf{k}} + \mathbf{v}_{n\mathbf{k}} \otimes \mathbf{v}_{n\mathbf{k}}$, and the subscripts $l$, $t_1$, and $t_2$ indicate properties belonging to the longitudinal and transverse modes.

Scattering by acoustic phonons through the piezoelectric interaction (``pi'') occurs in non-centrosymmetric systems and can dominate at low temperatures ($\lesssim$ \SI{50}{\kelvin}).
We have applied a similar treatment to extend the isotropic matrix element of \citet{meijer1953note}, \citet{harrison1956mobility}, and \citet{zook1964PiezoelectricScattering}, to include the full piezoelectric stress tensor $\mathbf{h}$ and scattering from all three acoustic modes.
The resulting matrix element is given by
\begin{widetext}
\begin{equation}
g_{nm}^\mathrm{pi}(\mathbf{k}, \mathbf{q}) =
   \sqrt{k_\mathrm{B} T} \sum_{\mathbf{G} \neq -\mathbf{q}}  \left[ 
        \frac{\mathbf{\hat{n}} \mathbf{h} \mathbin{:} \mathbf{\hat{S}}_l}{c_l\sqrt{\rho}} + 
        \frac{\mathbf{\hat{n}} \mathbf{h} \mathbin{:} \mathbf{\hat{S}}_{t_1}}{c_{t_1}\sqrt{\rho}} + 
        \frac{\mathbf{\hat{n}} \mathbf{h} \mathbin{:} \mathbf{\hat{S}}_{t_2}}{c_{t_2}\sqrt{\rho}}
    \right ]
    \frac{\mel{m\mathbf{k}+\mathbf{q}}{e^{i(\mathbf{q} + \mathbf{G})\cdot\mathbf{r}}}{n\mathbf{k}}}
         {\left | \mathbf{q} + \mathbf{G} \right |},
\label{eqn:element_pi}
\end{equation}
\end{widetext}
where $\mathbf{\hat{n}} = (\mathbf{q} + \mathbf{G}) / \left | \mathbf{q} + \mathbf{G} \right |$ is a unit vector in the direction of scattering. 
Due to the small energies of long-wavelength acoustic phonons, both piezoelectric and acoustic deformation potential scattering describe a purely elastic process.

We treat polar optical phonon scattering (``po'') by extending the Fr\"olich model  \cite{frohlich1954electrons} to include quantum mechanical wave function overlaps and anisotropic permittivity.
Here, electrons in a dielectric medium are perturbed by a dispersionless longitudinal optical phonon mode with frequency $\omega_\mathrm{po}$.
Our inelastic electron-phonon matrix element takes the form
\begin{equation}
\begin{aligned}
g_{nm}^\mathrm{po}(\mathbf{k}, \mathbf{q}) = &{}
    \left [ \frac{\hbar \omega_\mathrm{po}}{2} \right ] ^ {1/2} 
    \sum_{\mathbf{G} \neq -\mathbf{q}}
        \left (\frac{1}{\mathbf{\hat{n}}\cdot\boldsymbol{\epsilon}_\infty\cdot\mathbf{\hat{n}}} - \frac{1}{\mathbf{\hat{n}}\cdot\boldsymbol{\epsilon}_\mathrm{s}\cdot\mathbf{\hat{n}}}\right)
         ^\frac{1}{2} \\
    &{} \times \frac{\mel{m\mathbf{k}+\mathbf{q}}{e^{i(\mathbf{q} + \mathbf{G})\cdot\mathbf{r}}}{n\mathbf{k}}}
         {\left | \mathbf{q} + \mathbf{G} \right |},
\label{eqn:element_po}
\end{aligned}
\end{equation}
where  $\boldsymbol{\epsilon}_\mathrm{s}$ and $\boldsymbol{\epsilon}_\infty$ are the static and high-frequency dielectric tensors.
To capture scattering from the full phonon band structure in a single phonon frequency, each phonon mode is weighted by the dipole moment it produces (see Sec.~III(A) of the Supplemental Material) in line with recent work that has rederived the Fr\"olich model for systems with multiple phonon branches \cite{verdi2015frohlich,sjakste2015wannier}.
Both our extension of the Fr\"olich model and state-of-the-art first principles approaches produce similar matrix elements in the long-wavelength limit that dominates scattering (due to the polar singularity at $\mathbf{q}\rightarrow0$ \cite{verdi2015frohlich}).

Following the classic treatment of Brooks and Herring \cite{brooks1951scattering,herring1956transport} we consider the scattering from fully-ionized impurities  (`ii'')  modelled as screened Coulomb potentials, with the matrix element given by
\begin{equation}
g_{nm}^\mathrm{ii}(\mathbf{k}, \mathbf{q}) =
    \sum_{\mathbf{G} \neq -\mathbf{q}} 
     \frac{n_\mathrm{ii}^{1/2} Z e }{\mathbf{\hat{n}} \cdot \boldsymbol{\epsilon}_\mathrm{s} \cdot \mathbf{\hat{n}}}
    \frac{\mel{m\mathbf{k}+\mathbf{q}}{e^{i(\mathbf{q} + \mathbf{G})\cdot\mathbf{r}}}{n\mathbf{k}}}
         {\left | \mathbf{q} + \mathbf{G} \right | ^2 + \beta^2},
\label{eqn:element_ii}
\end{equation}
where $Z$ is the charge state of the impurity center, $e$ is the electron charge, $n_\mathrm{ii} = n_\mathrm{h} + n_\mathrm{e}$ is the concentration of ionized impurities, and $\beta$ is the inverse screening length (defined in Sec.~I(B) of the Supplemental Material).
Unlike previous formulations, our matrix element accounts for anisotropy in the charge screening through use of the full dielectric tensor.
Taken together, Eqs.~(\ref{eqn:rate}) and (\ref{eqn:element_ii}) reveal that the scattering almost diverges at long wavelengths ($\mathbf{q}\rightarrow0$) due to a $1/\abs{\mathbf{q}}^4$ dependence, and therefore requires very fine sampling to describe correctly.
For this reason, even the most sophisticated methods for calculating electron scattering by ionized impurities employ the Brooks--Herring formula, in which Eq.~(\ref{eqn:element_ii}) is analytically integrated for a single parabolic band \cite{chattopadhyay1981electron,ponce2018towards}.
To overcome this limitation, we employ a modified linear-tetrahedron approach to integration, in which tetrahedron cross sections are numerically resampled with hundreds of extra points that exactly satisfy the delta term in Eq.~(\ref{eqn:rate}).
This allows for ``effective'' $\mathbf{k}$-point mesh densities that would be almost impossible to achieve with uniform $\mathbf{k}$-point sampling (the full methodology is provided in Sec.~II(A) of the Supplemental Material).
Our approach enables, for the first time, evaluation of Coulomb based impurity scattering in systems with multiple non-parabolic bands, which even more sophisticated approaches do not currently implement.
In Sec.~II(E) of the Supplemental Material, we demonstrate that our methodology reproduces the exact Brooks--Herring mobility for parabolic band structures and reveal the failure of the Brooks--Herring approach in the case of systems containing multiple anisotropic valleys.

\begin{figure}
\includegraphics[width=0.85\linewidth]{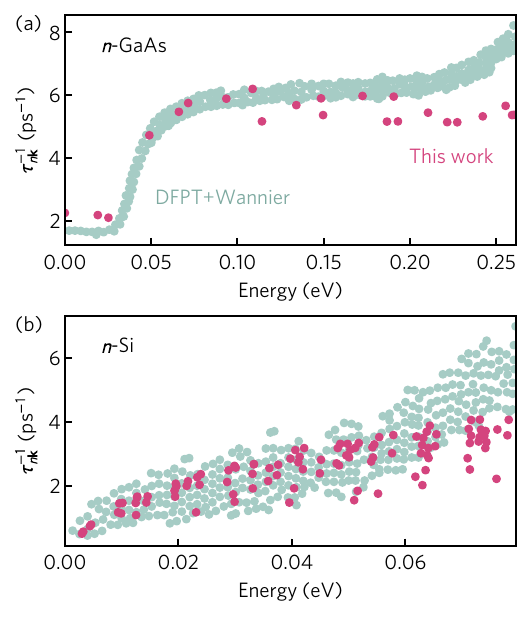}
\caption{\label{fig:rates}Comparison of the calculated scattering rates (pink) against those obtained using density functional perturbation theory combined with Wannier interpolation (light teal) for (a) $n$-GaAs 
\cite{zhou2016InitioElectron} and (b) $n$-Si \cite{ponce2018towards} at \SI{300}{\kelvin}.}
\end{figure}

\begin{figure}
\includegraphics[width=\linewidth]{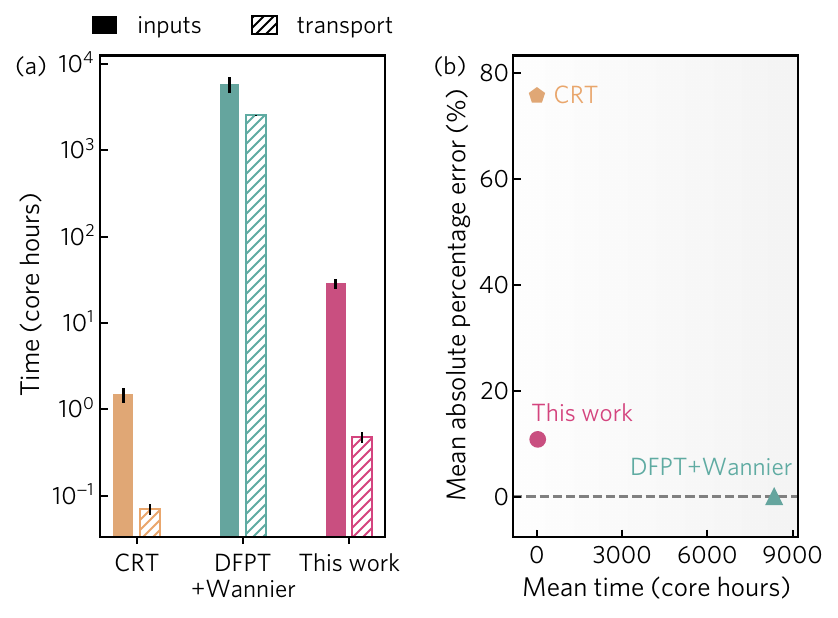}
\caption{\label{fig:timing}Existing methods for calculating electron transport properties are either computationally efficient but inaccurate (constant relaxation time, CRT) or accurate but highly computationally demanding (density functional perturbation theory combined with Wannier interpolation, DFPT+Wannier). The approach outlined in this work demonstrates accuracy comparable to state-of-the-art methods at approximately 1/500th the computational cost. (a) The time required to obtain electron mobility for each method is broken down by the time spent computing first-principles inputs and performing the scattering and transport calculations. (b) The mean absolute percentage error in the calculated mobility at \SI{300}{\kelvin} is compared to the total computational time (including the time to obtain all first-principles inputs). Results are averaged for NbFeSb ($p$-type, $n =$ \SI{2E20}{\per\cubic\centi\meter}, DFPT+Wannier \cite{samsonidze2018accelerated,zhou2018LargeThermoelectric}) and Ba\textsubscript{2}BiAu ($n$-type, $n =$ \SI{1E14}{\per\cubic\centi\meter}, DFPT+Wannier \cite{park2019HighThermoelectric}). In (b), the mobility error is referenced with respect to state-of-the-art DFPT+Wannier calculations as high-quality experimental data was not available. The full timing breakdown for each material is provided in Tables II and III in the Supplementary Material.}
\end{figure}

\begin{figure*}[t]
\includegraphics[width=\textwidth]{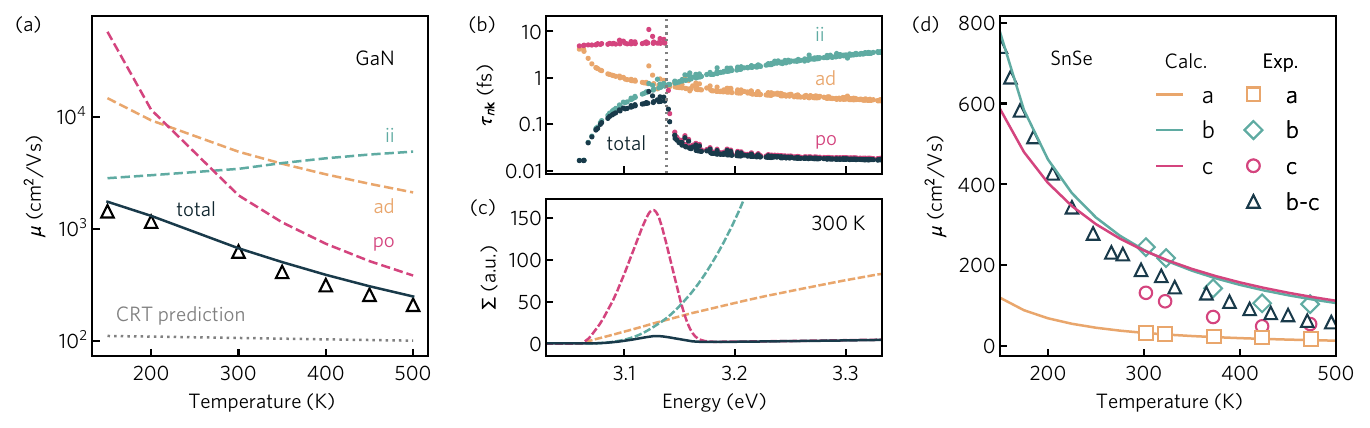}
\caption{\label{fig:material-results} (a) Comparison of the electron mobility of GaN against experiment (black triangles, \cite{sulkowski2010TransportProperties}). Mobility limited by ionized impurity (teal, ``ii''), acoustic deformation potential (orange, ``ad''), and polar optical phonon scattering (pink, ``po'') is indicated in dashed lines. Total  mobility taking into account all scattering mechanisms ($1 / \tau_{n\mathbf{k}}^\mathrm{ii} + 1 / \tau_{n\mathbf{k}}^\mathrm{ad} + 1 / \tau_{n\mathbf{k}}^\mathrm{po}$) is given by the black solid line. Constant relaxation time (CRT) calculations with $\tau =$ \SI{0.1}{\femto\second} is given by dotted gray line. (b) Electron lifetimes and (c) spectral conductivity arising from different scattering processes in GaN at \SI{300}{\kelvin}. The valence band maximum is set to zero \si{\electronvolt}. In (b), the vertical dotted gray line indicates the energy of the effective polar phonon frequency, $\omega_\mathrm{po}$. (d) Comparison of the direction-dependent mobility of SnSe against experiments --- $a$, $b$, $c$ points from Ref.~\cite{zhao2014UltralowThermal}, $b$-$c$ points from Ref.~\cite{asanabe1959ElectricalProperties}.}
\end{figure*}

The final $\mathbf{k}$-dependent scattering rates are obtained by integrating Eq.~(\ref{eqn:rate}) over all phonon wave vectors ($\mathbf{q}$) in the first Brillouin zone.
Elastic scattering processes are well described by the momentum relaxation time approximation (MRTA) to the BTE due to the requirement that $\tau_{n\mathbf{k}\rightarrow m\mathbf{k}+\mathbf{q}} = \tau_{m\mathbf{k}+\mathbf{q}\rightarrow n\mathbf{k}}$ \cite{ponce2020review}.
As this condition does not hold for inelastic processes, we adopt the self-energy relaxation time approximation (SERTA) to obtain the final polar phonon coupling rates \cite{ponce2018towards}.
Further justification for this approach is detailed in Sec.~I(A) of the Supplemental Material.
Electronic eigenvalues and group velocities needed to calculate scattering and transport properties are Fourier interpolated onto dense Brillouin zone grids using the \textsc{boltztrap2} software \cite{madsen2018boltztrap2} (as detailed in Sec.~II(A) of the Supplemental Material).
Electron mobility and Seebeck coefficient are calculated using the linearized Boltzmann transport equation via the Onsager transport coefficients \cite{onsager1931reciprocal,madsen2018boltztrap2} (see Sec.~I(C) of the Supplemental Material). We also employ a custom procedure for selecting the most important $\mathbf{k}$-points at which to calculate scattering to further reduce the computational expense (detailed in Sec.~II(B) of the Supplemental Material).

Unlike other state-of-the-art approaches in which a computationally expensive DFPT calculation is required to obtain $g(\mathbf{k}, \mathbf{q})$, in our method all matrix elements depend only on common materials parameters ($\omega_\mathrm{po}$, $\epsilon_\mathrm{s}$, $\epsilon_\mathrm{\infty}$, etc.) that can be calculated relatively inexpensively. Crucially, many of these properties are already tabulated in databases such as the Materials Project \cite{jain2013commentary} or can be obtained through relatively cheap \textit{ab initio} calculations. Furthermore, the matrix elements can be evaluated in a fixed time regardless of the number of atoms in the system, and multiple temperatures and carrier concentrations can be calculated simultaneously with only a modest increase in the computational time. Full timing information for the calculation of all first-principles inputs required to compute the transport properties of the materials discussed in this work and the scaling performance of each code routine is given in Sec.~II(D) of the Supplemental Material.

\subsection{Analysis of scattering rates and electron mobility}

In Figure (\ref{fig:rates}), we compare mode-dependent scattering rates for $n$-Si and $n$-GaAs calculated by our method against fully-first principles calculations (DFPT+Wannier) at \SI{300}{\kelvin} obtained using the \textsc{epw} and \textsc{perturbo} softwares \cite{ponce2018towards,zhou2020perturbo}.
The scattering of electrons in Si is dominated by acoustic phonons whereas polar optical phonon scattering dominates in GaAs, as revealed by the mobility analysis in Sec.~IV(B) in the Supplemental Material.
Excellent agreement is seen for both systems, with the onset of polar optical emission scattering in GaAs well described by our calculations.
Additional comparisons against DFPT+Wannier scattering rates for 3C-SiC and $p$-SnSe are provided in Fig.~(S11) of the Supplemental Material.
In both cases, the shape and magnitude of the scattering rates is well reproduced, particularly at low energies, despite the simpler approach that does not involve an expensive DFPT calculation to obtain the matrix elements.

In Fig.~(\ref{fig:timing}), we compare the time taken to compute the transport properties of NbFeSb and \ce{Ba2BiAu} (the full timing breakdown is tabulated in Tables II and III of the Supplemental Material).
Taking into account the time required to compute all first-principles inputs and the electron mobility at a single temperature and carrier concentration, our method offers over a 2 order of magnitude speed up compared to DFPT+Wannier (an average of 29 core hours versus 8,350 core hours).
Considering only the time needed to obtain the scattering rates and transport properties (i.e., presuming all inputs have already been tabulated), our approach offers a 4 order of magnitude speed up [Fig.~(\ref{fig:timing}a)].
This can be exploited when performing calculations at multiple temperatures and carrier concentrations.
For example, calculating the mobility of \ce{Ba2BiAu} for 10 temperatures requires approximately 32,000 core hours using DFPT+Wannier compared to less than 35 core hours with our approach (\SI{95}{\percent} of which is required to calculate the first principles inputs).
Furthermore, we expect the relative cost advantage of our method to increase with system size as unlike in DFPT+Wannier the computational expense of the matrix elements does not depend on the number of atoms.
This reduction in computational time, combined with similar accuracy to DFPT+Wannier [within \SI{10}{\percent}, see Fig.~(\ref{fig:timing}b)], makes our approach amenable to the large scale calculation of electronic transport properties.

Figure (\ref{fig:material-results}a) plots the calculated mobility of GaN against experimental measurements, indicating very close agreement from \SIrange{150}{500}{\kelvin}.
As each scattering mechanism is treated with a separate matrix element, this allows the impact of individual scattering processes to be assessed.
At low temperatures, the mobility of GaN is limited by impurity scattering, with polar optical phonon scattering dominating above \SI{300}{\kelvin}, as illustrated by the dashed lines in Fig.~(\ref{fig:material-results}a).
The total mobility taking into account all scattering mechanisms reproduces the experimental mobility with very high agreement.
Further insight into the competing nature of the scattering mechanisms is provided by the energy dependence of the electron lifetimes and the resulting spectral conductivity, $\Sigma(\varepsilon) = v(\varepsilon)^2 \tau(\varepsilon) N(\varepsilon)$ where $N$ is the density of states and $v$ is the group velocity, computed  at \SI{300}{\kelvin} and an electron concentration of \SI{5.5e16}{\per\cubic\centi\meter} [Figs.~(\ref{fig:material-results}b) and (\ref{fig:material-results}c)].
Impurity scattering dominates at the conduction band edge but diminishes quickly as energy increases.
At energies above $\omega_\mathrm{po}$ of the band minimum (above the phonon emission threshold), polar-optical interactions are two orders of magnitude stronger than any other competing mechanism and act as the primary limiting factor for electron mobility, in agreement with the experimental findings of Ref.~\cite{steigerwald1997gan} and DFPT+Wannier calculations \cite{ponce2019HoleMobility}.
In contrast, the mobility calculated using a constant relaxation time of $\tau =$ \SI{0.1}{\femto\second} --- a value on the higher end of that typically employed in screening studies \cite{miyata2018HighThroughputScreening,madsen2006AutomatedSearch,bhattacharya2016NovelPtype} --- underestimates the mobility by a factor of 2--10 depending on the temperature, as shown in Fig.~(\ref{fig:material-results}a). More fundamentally, the CRTA does not reproduce the correct shape of temperature dependence as depicted in Fig.~(\ref{fig:material-results}c).
The ability of our method to reproduce the qualitative temperature dependence of transport properties, as well as make good approximations of quantitative behavior (often closely in-line with more detailed theoretical methods), thus represents a major advance for improving the accuracy of high-throughput methods.

A primary goal of the present approach is to extend well established scattering matrix elements that were formulated for isotropic materials properties to be compatible with highly anisotropic materials.
To that end, we have calculated the direction-dependent hole mobilities of \textit{Pnma} structured SnSe at a carrier concentration of \SI{3e17}{\per\cubic\centi\meter}, with the results compared to Hall measurements in Fig.~(\ref{fig:material-results}d).
Single-crystal SnSe has recently attracted significant attention as a thermoelectric material.
Due to its layered structure, SnSe exhibits anisotropic transport properties, with the highest thermoelectric performance observed along the $b$ axis \cite{zhao2014UltralowThermal}.
Our calculations reproduce the strong directional dependence in transport measurements, in which the mobility parallel to the layers (along $b$ and $c$) is almost an order of magnitude larger than that perpendicular to the layers (along $a$).
Our mobility results agree remarkably well with the considerably more computationally expensive electron-phonon calculations performed using DFPT+Wannier and $G0W0$ band structures \cite{ma2018IntrinsicPhononlimited} (see Fig.~(S6) in the Supplemental Material).
We note that additional anisotropy in the mobility between the $b$ and $c$ directions has been observed in high temperature experimental measurements \cite{zhao2014UltralowThermal}.
In both our calculations and DFPT+Wannier, however, the mobility along $b$ and $c$ are almost the same for temperatures above \SI{300}{\kelvin} \cite{ma2018IntrinsicPhononlimited}.
The discrepancy against experiment is thought to derive from the use of a Hall factor $r_H$ of unity when extracting the carrier concentrations needed to compute mobility \cite{ma2018IntrinsicPhononlimited}.

\begin{figure}
\includegraphics[width=0.95\linewidth]{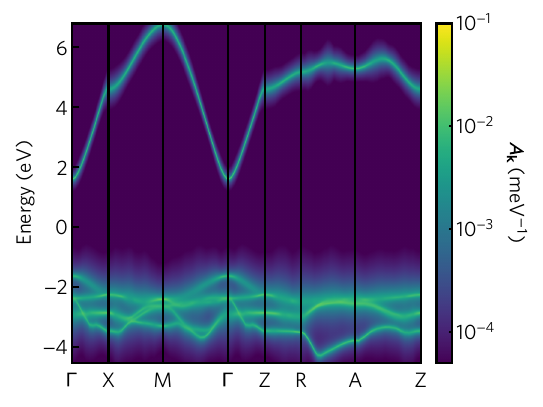}
\caption{\label{fig:lineshape} Spectral band structure of SnO$_2$ indicating band and $\mathbf{k}$-dependent electron linewidths calculated at \SI{300}{\kelvin}.}
\end{figure}

Access to band and $\mathbf{k}$-dependent lifetimes can further be used to calculate electron linewidths that are qualitatively comparable to those measured through techniques such as angle-resolved photoemission spectroscopy (ARPES) \cite{park2008VanHove}.
In Fig.~(\ref{fig:lineshape}) we plot the spectral band structure of \ce{SnO2} along a high symmetry Brillouin zone path, where the spectral function 
$A_\mathbf{k}(\mathbf{\varepsilon}) = \pi^{-1} \sum_n (\tau_{n\mathbf{k}}^{-1}/2) / [(\varepsilon - \varepsilon_{n\mathbf{k}} / \hbar)^2 +(\tau_{n\mathbf{k}}^{-1}/2)^{2}]$ 
was calculated at \SI{300}{\kelvin}.
The spectral function provides insight into the $\mathbf{k}$-dependence of the carrier lifetimes.
States close to the conduction band edge at $\Gamma$ exhibit long lifetimes (low energy broadening) due to the reduced phase space of available states for scattering.
Between the Z and R high symmetry points, the lowest conduction band is relatively flat leading to large scattering rates and considerable broadening of the spectral function.

\subsection{Electron mobility and Seebeck coefficient across many systems}

\begin{figure}
\includegraphics[width=0.9\linewidth]{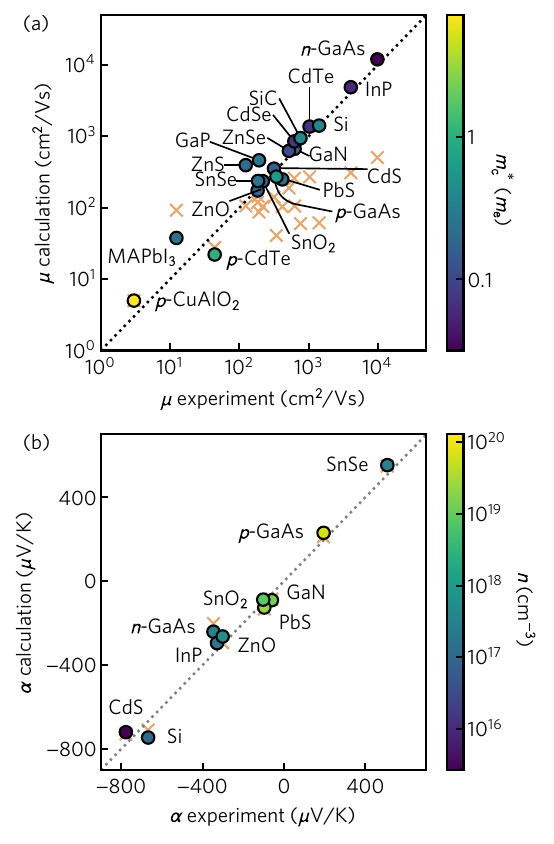}
\caption{\label{fig:screening-results}(a) Comparison of carrier mobilities at \SI{300}{\kelvin} between calculations and experiments, with points colored by the conductivity effective mass $m^*_c$.~(b) Comparison of Seebeck coefficients at \SI{300}{\kelvin} between calculations and experiments, with points colored by the majority carrier concentration $n$. For Si and CdS we compare directly to the diffusive component of Seebeck coefficient only. CH$_3$NH$_3$PbI$_3$ has been abbreviated as MAPbI$_3$.
In (a) and (b), orange crosses indicate results computed using a constant relaxation time of \SI{0.1}{\femto\second}. Detailed temperature and carrier concentration results for each material are provided in Sec.~IV of the Supplemental Material.}
\end{figure}

To demonstrate the generality of our approach, we investigate the transport properties of 17 semiconductors ranging from 2 to 48 atoms in their primitive unit cells.
To highlight the compatibility of the method with high-throughput computations, all inputs (eigenvalues, wave functions, materials parameters) are obtained from density functional theory (DFT) using low cost exchange--correlation functionals (see Methods section).
All such materials parameters are listed in Table IV of in Supplemental Material.
Results are compared to transport measurements on high purity single-crystalline samples to minimize the effects of grain boundaries and crystallographic defects. Further details on the calculation methodology and selection of reference data are provided in Secs.~II and III in the Supplemental Material.
The materials span multiple chemistries, doping polarities, and band structure types including anisotropic and multiband systems, and comprise:~(i) conventional semiconductors, Si, GaAs, GaN, GaP, InP, ZnS, ZnSe, CdS, CdSe, and SiC;~(ii) the thermoelectric candidate SnSe;~(iv) photovoltaic absorbers PbS and CdTe;~and (iii) transparent conductors, \ce{SnO2}, ZnO, and \ce{CuAlO2}. Our dataset also includes the relatively complex \ce{CH3NH3PbI3} hybrid perovskite containing 48-atoms.
In Figure (\ref{fig:screening-results}a) we compare calculated mobility against experimental measurements for all 17 materials in our dataset.
Calculations were performed using the experimentally determined carrier concentrations at a temperature of \SI{300}{\kelvin}.
Results regarding the temperature and carrier concentration dependence of mobility for all materials (calculated, experimental, and comparison with CRTA) is provided in Figs.~(S6) and (S7) of the Supplemental Material and include the breakdown of mobility by scattering type. These plots represent a comprehensive test of AMSET, across many materials, not only at the single condition plotted in Figure (\ref{fig:screening-results}a) but when conditions are varied.

The calculated mobilities agree closely with experiment across all materials, covering several orders of magnitude from ZnO (180$\,$cm$^{2}$/Vs) to $n$-type GaAs ($\mu_\mathrm{exp} =$ $2.1\times10^{4}\,$cm$^{2}$/Vs).
Notably, the calculated mobility (Spearman rank coefficient against experiment $r_\mathrm{s} = $ 0.92) improves significantly on results obtained using a constant relaxation time approximation ($r_\mathrm{s} = $ 0.50).
As we demonstrate in Fig.~(S12) of the Supplemental Material, our method also improves the dependence of mobility on temperature, obtaining a mean squared error (MSE) of 0.21 consistent with DFPT+Wannier (MSE = 0.20) and dramatically more accurate than a constant relaxation time (MSE = 3.2).
Greater deviation from experiment is observed for materials with smaller mobilities such as $p$-\ce{CuAlO2} (3\,cm$^{2}$/Vs in $a$-$b$ plane), where a local hopping mechanism is proposed to compete with band transport \cite{kawazoe1997cualo2}, and $p$-CdTe in which spin--orbit coupling (SOC) is known to dramatically impact the scattering rates at the valence band edge \cite{yadav2019charge} but was not included in our calculations.
Additional deviation is observed for $n$-ZnS, where the calculated mobility is almost a factor of 4 larger than Hall measurements.
We find this overestimation is largely due to the underestimation of the conduction band effective mass arising from use of the PBE exchange--correlation functional ($m^{*,\mathrm{PBE}}_c =$ 0.16$\,m_e$) when compared to experiment ($m^{*,\mathrm{exp}}_c =$ 0.22$\,m_e$) \cite{nagano1991PhotoluminescenceSbDoped}.
As we detail in Fig.~(S8) of the Supplemental Material, calculations performed using the hybrid HSE06 functional result in a larger effective mass ($m^{*,\mathrm{HSE}}_c =$ 0.20$\,m_e$) and improved agreement with the experimental mobility.
Lastly, the deviation seen for \ce{CH3NH3PbI3} is likely due to the use of polycrystalline thin films in experimental measurements.
As highlighted by Fig.~(S6) of the Supplemental Material, our temperature-dependent results are in excellent agreement (within \SI{5}{\percent} at all temperatures) against fully-first principles calculations performed using \textsc{epw} \cite{ponce2019OriginLow}.
The ability of our approach to accurately describe the electron-phonon coupling of a highly complex structure with 144 phonon-modes while remaining computationally efficient
highlights its potential in high-throughput screening of transport behaviour.

Accurate calculation of Seebeck coefficients is of primary interest in the prediction and analysis of thermoelectric materials.
In Fig.~(\ref{fig:screening-results}b) we compare calculated Seebeck coefficients against those obtained experimentally at \SI{300}{\kelvin}.
A comparison of the temperature dependence of the Seebeck coefficient for all materials is provided in Fig.~(S9) of the Supplemental Material.
We see reasonable agreement against experiment across the full range of materials, for both $p$- and $n$-type samples, corresponding to positive and negative Seebeck coefficients, respectively.
In the Fig.~S10 of the Supplemental Material we demonstrate that use of the HSE06 hybrid functional can further improve the agreement against experiment for $n$-type GaAs.
We note that for Si and CdS we compare directly to the diffusive component of Seebeck coefficient only, ignoring the effects of phonon drag which contribute substantially even at room temperature \cite{herring1954theory,geballe1955seebeck,fiorentini2016thermoelectric,morikawa1965SeebeckEffect}.
The Seebeck coefficient displays a weaker dependence on electron lifetimes than mobility and conductivity and so is often treated within the CRTA (in which case the specific relaxation time cancels in the equations). \cite{ricci2017ab}.
Figure (\ref{fig:screening-results}b) indicates that this approximation is often justified due to the relatively small disagreements between constant relaxation time and mode-dependent relaxation time results, in-line with previous comparisons of CRTA against experimental data
 \cite{chen2016understanding}. 

\section{Discussion}

A key motivation in the development of the present approach is the opportunity to obtain accurate carrier lifetimes at minimal computational expense.
Ideally, the method should be cheap enough to permit the calculation of transport properties for thousands of compounds in a high-throughput manner as well as large and complex materials. This would allow for reliable screening of materials for functional applications as well as enable investigations of systems with larger unit cells and more complex crystal structures.
We stress that an expensive DFPT calculation is not required to obtain the matrix elements unlike methods such as \textsc{epw} \cite{ponce2016epw}, \textsc{perturbo} \cite{zhou2020perturbo}, and \textsc{epic star} \cite{deng2020EPICSTAR}.
In our approach, the primary computational expense is the calculation of first-principles inputs, particularly the dielectric constant as detailed in Table II of the Supplemental Material.
However, due to our use of the relatively low-cost PBE exchange--correlation functional all inputs (electronic structure, $\Gamma$-point phonon frequencies, elastic constants, dielectric constants and piezoelectric tensor) can be obtained with moderate computational requirements (generally less than 50 core hours to compute all properties, see Table II of the Supplemental Material).
The calculation of transport properties takes even less time; the results for each material presented in this work were computed in under an hour on a personal laptop --- further timing analysis, indicating the breakdown for different routines in the code, is presented in Fig.~(S4) of the Supplemental Material.
In addition, many of the materials properties required to calculate the scattering matrix elements are already available in computational materials databases.
For example, at the time of writing the Materials Project contains over 3,300 piezoelectric tensors, 7,100 dielectric constants and phonon frequencies, and over 13,000 elastic constants \cite{jain2013commentary,jong2015database,dejongChartingCompleteElastic2015}.
Accordingly, our approach is well suited for the large-scale analysis of transport properties.
To that end, we have made available a Python implementation of the method called \textit{Ab initio} Scattering and Transport (\textsc{amset}) at \url{https://github.com/hackingmaterials/amset}.
Our goal is for this software to complement higher level methods, such as \textsc{epw} \cite{ponce2016epw} and \textsc{perturbo} \cite{zhou2020perturbo}, which are state-of-the-art but considerably more computationally demanding.
A schematic overview of the package, indicating the inputs, outputs and command-line tools is given in Fig.~(S3) of the Supplemental Material.

We stress that all electronic dispersions and wave functions were computed using the PBE functional which tends to over-delocalise electronic states and underestimate effective masses.
In most cases, the calculated mobility is overestimated compared to experiment, suggesting that use of higher level methods such as hybrid DFT or GW will be beneficial.
In addition, there are several limitations of the current approach that may be addressed in a future release. In particular, optical deformation potential scattering is not treated, the symmetry of phonon modes is not used for filtering scattering events, and our matrix elements are not yet suitable for metals.

In conclusion, we introduced a method for calculating electron lifetimes and transport properties of semiconductors and insulators.
Our method extends isotropic scattering matrix elements to support highly anisotropic materials and relies on a novel approach to Brillouin zone integration that overcomes the need for extremely dense $\mathbf{k}$-point sampling.
The present approach offers similar accuracy to state-of-the art methods at approximately 1/500th the computational cost and relies only on inputs that can be obtained from low-cost \textit{ab initio} methods and that are routinely available in computational materials science databases.
Furthermore, the agreement of mobility against experiment (Spearman rank coefficient $r_\mathrm{s} = $ 0.92) improves significantly on other low-cost methods such as a constant relaxation time approximation ($r_\mathrm{s} = $ 0.50) and temperature dependence is accurately captured. We expect that our method will enable accurate screening of transport properties in high-throughput computational workflows.

\section{Methods}

All DFT calculations were performed using the Perdew--Burke--Ernzerhof (PBE) exchange--correlation functional \cite{perdew1996generalized} as implemented in the Vienna \textit{ab initio} Simulation Package (\textsc{vasp}) \cite{kresse1996efficient,kresse1996efficiency}.
Materials parameters, including elastic constants, dielectric tensors, deformation potentials, and phonon frequencies are listed in Table IV of in Supplemental Material.
Calculations were performed in a plane-wave basis set with scalar relativistic psueodpoentials and with the interactions between core and valence electrons described using the projector augmented-wave method (PAW) \cite{blochlProjectorAugmentedwaveMethod1994,Kresse1999}.
The set-up, submission, and management of first-principles calculations were handled using the \textsc{atomate} workflow management software with the default parameters of version \verb|0.8.3| \cite{mathewAtomateHighlevelInterface2017,AtomateV02018}.
The plane-wave energy cutoff was set to \SI{520}{\electronvolt}.
Structure optimization was performed with a reciprocal $\mathbf{k}$-point density of \SI{64}{\textbf{k}\mbox{-}points\per\cubic\angstrom}.
The uniform non-self-consistent calculations used as input to the scattering calculations were run with a reciprocal $\mathbf{k}$-point density of \SI{1000}{\textbf{k}\mbox{-}points\per\cubic\angstrom}.
Band gaps are corrected using a scissor operation to match those calculated by the Heyd--Scuseria--Ernzerhof (HSE06) hybrid functional \cite{heyd2003hybrid,heyd2006erratum}.
Piezeoelectric constants, and static and high-frequency dielectric constants were computed using density functional perturbation theory (DFPT) based on the method developed and by \citet{baroni1986InitioCalculation} and adapted to the PAW formalism by \citet{Gajdos2006}.
Elastic constants were obtained through the stress-strain approach detailed in Ref.~\cite{dejongChartingCompleteElastic2015}.
Spin--orbit interactions were included for calculations on \ce{CH3NH3PbI3} as they were necessary to obtain the correct band ordering at the conduction band minimum.
A comparison of the experimental and HSE06 band gaps, along with initial and interpolated $\mathbf{k}$-point meshes are provided in Table V of the Supplemental Material.
All timing information (first-principles inputs and transport properties) displayed in Fig.~(\ref{fig:timing}a) was obtained on an Intel Xeon Haswell processor E5-2698 v3 @ 2.3 GHz, except the EPW timing for NbFeSb.
EPW timing information for NbFeSb was reported in Ref.~\cite{samsonidze2018accelerated} without specifying the processor type, so we have assumed a 1:1 correspondence in core performance.

\begin{acknowledgements}

This work was intellectually led and funded by the U.S. Department of Energy (DOE), Office of Basic Energy Sciences, Early Career Research Program. This research used resources of the National Energy Research Scientific Computing Center, which is supported by the Office of Science of the U.S. Department of Energy under Contract No. DEAC02-05CH11231. Lawrence Berkeley National Laboratory is funded by the DOE under award DE-AC02-05CH11231.
We acknowledge fruitful discussions with K. Inzani and T. Karin.

\end{acknowledgements}

\providecommand{\noopsort}[1]{}\providecommand{\singleletter}[1]{#1}%
%


\end{document}


\preprint{APS/123-QED}

\title{Efficient calculation of electron scattering rates from first principles}

\author{Alex M. Ganose}
\affiliation{Energy Technologies Area, Lawrence Berkeley National Laboratory, Berkeley, California 94720, USA}

\author{Junsoo Park}
\affiliation{Energy Technologies Area, Lawrence Berkeley National Laboratory, Berkeley, California 94720, USA}

\author{Alireza Faghaninia}
\affiliation{Energy Technologies Area, Lawrence Berkeley National Laboratory, Berkeley, California 94720, USA}

\author{Rachel Woods-Robinson}
\affiliation{Energy Technologies Area, Lawrence Berkeley National Laboratory, Berkeley, California 94720, USA}
\affiliation{Department of Materials Science and Engineering, University of California Berkeley, California 94720, United States}

\author{Kristin A. Persson}
\affiliation{Department of Materials Science and Engineering, University of California Berkeley, California 94720, United States}
\affiliation{Molecular Foundry, Energy Sciences Area,  Lawrence Berkeley National Laboratory, Berkeley, California 94720, USA}

\author{Anubhav Jain}
\affiliation{Energy Technologies Area, Lawrence Berkeley National Laboratory, Berkeley, California 94720, USA}

\date{\today}

\maketitle

\section{Theoretical Framework}

\subsection{Linearized Boltzmann transport equation}

Electron mobility, $\mu_e$, can be computed through the linearized Boltzmann transport equation (BTE) \cite{ziman1960electrons,ponce2020review,ponce2018towards,li2015ElectricalTransport}, given for electrons as
\begin{equation}
    \mu_{e,\alpha\beta} = \frac{-1}{n_e \Omega} 
    \sum_{n\in\mathrm{cb}} \int \frac{\dd{\mathbf{k}}}{\Omega_\mathrm{BZ}}
    v_{n\mathbf{k},\alpha}\partial_{E_\beta} f_{n\mathbf{k}},
    \label{eq:bte}
\end{equation}
where $\alpha$ and $\beta$ denote Cartesian coordinates, $n_e$ is the electron concentration, $\Omega$ and $\Omega_\mathrm{BZ}$ are the volumes of the unit cell and first Brillouin zone, respectively, $v_{n\mathbf{k},\alpha}$ is the group velocity of band index $n$ and wave vector $\mathbf{k}$, ``cb'' stands for conduction bands, and $\partial_{E_\beta} f_{n\mathbf{k}}$ is the perturbation to the Fermi--Dirac distribution by an electric field $\mathbf{E}$. The Fermi--Dirac distribution is given by
\begin{equation}
    f^0_{n\mathbf{k}} = \frac{1}{\exp\left[(\varepsilon_{n\mathbf{k}}-\varepsilon_\mathrm{F})/k_\mathrm{B}T\right] + 1},
\end{equation}
where $\varepsilon_{n\mathbf{k}}$ is the energy of state $n\mathbf{k}$, $\varepsilon_\mathrm{F}$ is the Fermi level, $k_\mathrm{B}$ is the Boltzmann constant, and $T$ is temperature.
The perturbation to the equilibrium Fermi--Dirac distribution is given by the self-consistent solution of
\begin{equation}
\begin{aligned}
    \partial_{E_\beta} f_{n\mathbf{k}} ={} & e \pdv{f_{n\mathbf{k}}^0}{\varepsilon_{n\mathbf{k}}} 
    v_{n\mathbf{k},\beta}\tau_{n\mathbf{k}}
    + \frac{2 \pi \tau_{n\mathbf{k}}}{\hbar}
    \sum_{m} \int \frac{\dd{\mathbf{q}}}{\Omega_\mathrm{BZ}}
    \lvert g_{nm}(\mathbf{k}, \mathbf{q}) \rvert^2 \\
        {}& \times [ (n_\mathbf{q} + 1 - f^0_{n\mathbf{k}})
        \delta(\Delta\varepsilon_{\mathbf{k},\mathbf{q}}^{nm} + \hbar\omega_\mathbf{q}) \\
        {}& + (n_\mathbf{q} + f^0_{n\mathbf{k}})
        \delta(\Delta\varepsilon_{\mathbf{k},\mathbf{q}}^{nm} - \hbar\omega_\mathbf{q})] \partial_{E_\beta} f_{m\mathbf{k}+\mathbf{q}},
\label{eq:perturbation}
\end{aligned}
\end{equation}
where  $\tau_{n\mathbf{k}}$ is the electron lifetime, $\delta$ is the Dirac delta function, $\Delta\varepsilon_{\mathbf{k},\mathbf{q}}^{nm} = \varepsilon_{n\mathbf{k}} -  \varepsilon_{m\mathbf{k} + \mathbf{q}}$, $\hbar$ is the reduced Planck constant, and $n_\mathbf{q}$ is the Bose--Einstein occupation.
The  matrix elements $g_{nm}\left(\mathbf{k}, \mathbf{q} \right )$ give the probability of scattering from an initial state $n\mathbf{k}$ to final state $m\mathbf{k} + \mathbf{q}$ via a phonon with wave vector $\mathbf{q}$ and frequency $\omega_\mathbf{q}$.

The primary complexity in the Boltzmann transport equation results from the dependence of the linear response coefficients $\partial_{E_\beta} f_{n\mathbf{k}}$ of state $n\mathbf{k}$ on all other states $m\mathbf{k} + \mathbf{q}$.
Accordingly, there are several common approximations to the BTE that can significantly reduce the computational cost.
The \textit{momentum relaxation time approximation} (MRTA) makes two simplifications:~(i) Firstly, the linear response coefficients are presumed to only act in the direction of the band velocity, such that the electron lifetimes will be \textit{scalar} quantities \cite{li2015ElectricalTransport,ponce2020review}.~(ii) Secondly, the probability of scattering from state $n\mathbf{k}$ to $m\mathbf{k} + \mathbf{q}$ is assumed to be the same as scattering from state $m\mathbf{k} + \mathbf{q}$ to $n\mathbf{k}$.
The result is that the effects of back scattering are accounted for by a geometrical factor resulting from the electronic group velocities.
The resulting expression for $\tau_{n\mathbf{k}}^{-1}$ can be written
\begin{equation}
\begin{aligned}
    \tau_{n\mathbf{k}}^{-1} = \sum_m \int \frac{\dd{\mathbf{q}}}{\Omega_\mathrm{BZ}}
    \left [  1 -  \frac{\mathbf{v}_{n\mathbf{k}} \cdot \mathbf{v}_{m\mathbf{k} + \mathbf{q}}}{\abs{\mathbf{v}_{n\mathbf{k}}}^2} \right]
    \tau_{n\mathbf{k}\rightarrow m\mathbf{k}+\mathbf{q}}^{-1},
\end{aligned}
\label{eq:mrta-rate}
\end{equation}
where $\tau_{n\mathbf{k}\rightarrow m\mathbf{k}+\mathbf{q}}^{-1}$ is the partial decay rate for scattering from initial state $n\mathbf{k}$ to final state $m\mathbf{k} + \mathbf{q}$.
In this approximation, Eq.~(\ref{eq:bte}) can be rewritten
\begin{equation}
    \mu_{e,\alpha\beta}^\mathrm{MRTA} = \frac{e}{n_e \Omega} 
    \sum_{n\in\mathrm{cb}} \int \frac{\dd{\mathbf{k}}}{\Omega_\mathrm{BZ}}
    \pdv{f^0_{n\mathrm{k}}}{\varepsilon_{n\mathbf{k}}}
    v_{n\mathbf{k},\alpha} v_{n\mathbf{k},\beta}\tau_{n\mathbf{k}}.
    \label{eq:mob-mrta}
\end{equation}
A further simplification can be made by ignoring the effcts of scattering back into the state $n\mathbf{k}$ entirely. This corresponds to neglecting the second term on the right-hand side of Eq.~(\ref{eq:perturbation}) or setting the geometric factor in the square bracket of Eq.~(\ref{eq:mrta-rate}) to 1.
In this approach, termed the \textit{self-energy relaxation time approximation} (SERTA) \cite{ponce2018towards}, the electron lifetimes can be obtained according to
\begin{equation}
\begin{aligned}
    \tau_{n\mathbf{k}}^{-1} = \sum_m \int \frac{\dd{\mathbf{q}}}{\Omega_\mathrm{BZ}}
    \tau_{n\mathbf{k}\rightarrow m\mathbf{k}+\mathbf{q}}^{-1},
\end{aligned}
\label{eq:serta-rate}
\end{equation}
and the mobility calculated in the same manner as Eq.~(\ref{eq:mob-mrta}).

The partial decay rates of Eqs.~(\ref{eq:mrta-rate}) and (\ref{eq:serta-rate}) can be obtained through Fermi's golden rule.
In the present work, we implement two classes of scattering:~(i)
inelastic scattering which occurs via emission or absorption of a phonon and (ii) perfectly elastic scattering in which no energy is gained or lost.
In the case of inelastic scattering, the partial decay rate can be written \cite{grimvall1981electron,giustino2007electron}
\begin{equation}
\begin{aligned}
    \tau_{n\mathbf{k}\rightarrow m\mathbf{k}+\mathbf{q}}^{-1} ={}&
        \frac{2\pi}{\hbar}   \lvert g_{nm}(\mathbf{k}, \mathbf{q}) \rvert^2 \\
         &{} \times [ (n_\mathbf{q} + 1 - f^0_{m\mathbf{k} + \mathbf{q}})
        \delta(\Delta\varepsilon_{\mathbf{k},\mathbf{q}}^{nm} - \hbar\omega_\mathbf{q}) \\
        &{} + (n_\mathbf{q} + f^0_{m\mathbf{k} + \mathbf{q}})
        \delta(\Delta\varepsilon_{\mathbf{k},\mathbf{q}}^{nm} + \hbar\omega_\mathbf{q})],
\end{aligned}
\label{eq:inelastic-partial-rate}
\end{equation}
where the $-\hbar\omega_\mathbf{q}$ and $+\hbar\omega_\mathbf{q}$ terms correspond to scattering by emission and absorption of a phonon, respectively.
The dependence of $\tau_{n\mathbf{k}\rightarrow m\mathbf{k}+\mathbf{q}}^{-1}$ on the occupation of state $m\mathbf{k} + \mathbf{q}$ and the observation that $f_{m\mathbf{k} + \mathbf{q}} \neq f_{n\mathbf{k}}$ reveals that inelastic scattering is not commutative --- i.e., $\tau_{n\mathbf{k}\rightarrow m\mathbf{k}+\mathbf{q}}^{-1} \neq \tau_{m\mathbf{k}+\mathbf{q}\rightarrow n\mathbf{k}}^{-1}$. We note that for spin polarized materials, scattering only occurs between states in the same spin channel --- i.e., there are no interactions between spin-up and spin-down electrons.

For elastic scattering, Eq.~(\ref{eq:inelastic-partial-rate}) reduces to
\begin{equation}
\begin{aligned}
    \tau_{n\mathbf{k}\rightarrow m\mathbf{k}+\mathbf{q}}^{-1} =
        \frac{2\pi}{\hbar} &{}  \lvert g_{nm}(\mathbf{k}, \mathbf{q}) \rvert^2 \delta{\left ( \Delta\varepsilon_{\mathbf{k},\mathbf{q}}^{nm} \right )}.
\end{aligned}
\label{eq:elastic-partial-rate}
\end{equation} 
In contrast to inelastic scattering, elastic processes do not depend on the occupation of state $m\mathbf{k}+\mathbf{q}$.
Accordingly, $\tau_{n\mathbf{k}\rightarrow m\mathbf{k}+\mathbf{q}}^{-1} = \tau_{m\mathbf{k}+\mathbf{q}\rightarrow n\mathbf{k}}^{-1}$ and a primary assumption of the MRTA is satisfied.
For this reason, we treat elastic scattering processes under the MRTA, whereas inelastic scattering processes are treated in the SERTA.

\subsection{Scattering matrix elements}

\begin{table}[t]
\caption{Summary of scattering mechanisms}
\label{tab:matrix-elements}
\begin{ruledtabular}
\begin{tabular}{llll}
Name &
  Required properties &
  Type &
  Refs. \\ \colrule
Ionized impurity &
  Static dielectric &
  Elastic &
  \cite{brooks1951scattering,herring1956transport} \\[10pt]
  \begin{tabular}[c]{@{}l@{}}Acoustic \\ deformation \\ potential\end{tabular}
  &
  \begin{tabular}[c]{@{}l@{}}Deformation \\ potential, \\ elastic constant\end{tabular} &
  Elastic &
  \cite{bardeen1950deformation,khan1984DeformationPotentials,kartheuser1986DeformationPotentials,resta1991DeformationpotentialTheorem} \\[20pt]
\begin{tabular}[c]{@{}l@{}}Piezoelectric \\ acoustic \end{tabular} & \begin{tabular}[c]{@{}l@{}} Piezoelectric constant \end{tabular} &  Elastic &
  \cite{meijer1953note,harrison1956mobility,rode1975low} \\[15pt]
\begin{tabular}[c]{@{}l@{}}Polar optical \\ phonon\end{tabular} &
  \begin{tabular}[c]{@{}l@{}}Static and \\ high-frequency \\  dielectric, \\ phonon frequency\end{tabular} &
  Inelastic &
  \cite{frohlich1954electrons} \\[15pt]
\end{tabular}
\end{ruledtabular}
\end{table}

The general form of the quantum mechanical scattering matrix elements in Eqs.~(\ref{eq:perturbation}), (\ref{eq:mrta-rate}), and (\ref{eq:serta-rate}) is
\begin{equation}
    g_{nm}(\mathbf{k}, \mathbf{q}) = \mel{m\mathbf{k}+\mathbf{q}}{\Delta_\mathbf{q}V}{n\mathbf{k}}
\end{equation}
where $\Delta_\mathbf{q}V$ is an electronic perturbation associated with a scattering process \cite{giustino2007electron}.
In the present work we calculate matrix elements within the Born approximation \cite{born1926quantenmechanik};
namely, the electronic perturbation is assumed to only weakly impact the wave function of the final state  $m\mathbf{k}+\mathbf{q}$.
The scattering matrix elements considered in this work and the materials parameters needed to calculate them are summarized in Table \ref{tab:matrix-elements}.

\subsubsection{G-vector summation}

The matrix elements include a sum over reciprocal lattice vectors $\mathbf{G}$.
In this work, we restrict the summation to only include a single reciprocal lattice vector, $\mathbf{G}_\mathbf{q}$, such that $\left | \mathbf{q} + \mathbf{G}_\mathbf{q} \right | = \min_\mathbf{G}  \left | \mathbf{q} + \mathbf{G}_\mathbf{q} \right |$.
This corresponds to retaining only the handful of $\mathbf{G} + \mathbf{q}$ vectors that define the first Brillouin zone.

\subsubsection{Impurity scattering}

The inverse screening length $\beta$, required in the calculation of the ionized impurity matrix element, is given by
\begin{equation}
    \beta^2 =  \frac{e^2}{\epsilon_\mathrm{s}  k_\mathrm{B} T \Omega}
       \sum_n \int f^0_{n\mathbf{k}} (1 - f^0_{n\mathbf{k}}) \dd{\mathbf{k}},
\label{eq:inv-screening}
\end{equation}
where $1/\beta$ corresponds to the Debye length and Thomas--Fermi screening length for non-degenerate and degenerate doping regimes, respectively \cite{rode1971electron}.

\subsection{Transport properties}

Electronic transport properties --- namely, conductivity, Seebeck coefficient, and electronic component of thermal conductivity --- are calculated through the Onsager coefficients \cite{onsager1931reciprocal,madsen2018boltztrap2}.
The spectral conductivity, defined as
\begin{equation}
    \Sigma_{\alpha\beta}(\varepsilon) =  \sum_n \int \frac{\dd{\mathbf{k}}}{8\pi^3} 
    v_{n\mathbf{k},\alpha}v_{n\mathbf{k},\beta}\tau_{n\mathbf{k}}
    \delta{\left(\varepsilon - \varepsilon_{n\mathbf{k}} \right )},
    \label{eq:spectral-cond}
\end{equation}
is used to compute the moments of the generalized transport coefficients
\begin{equation}
    \mathcal{L}^n_{\alpha\beta} = e^2 \int \Sigma_{\alpha\beta}(\varepsilon) (\varepsilon_\mathrm{F} - \varepsilon)^n
    \left [ -\pdv{f^0}{\varepsilon} \right ] \dd{\varepsilon},
    \label{eq:transport-moments}
\end{equation}
where $\varepsilon_\mathrm{F}$ is the Fermi level at a certain doping concentration and temperature $T$.
Electrical conductivity ($\sigma$), Seebeck coefficient ($S$), and the charge carrier contribution to thermal conductivity ($\kappa$) are obtained as
\begin{align}
    \sigma_{\alpha\beta} ={}& \mathcal{L}_{\alpha\beta}^0, \\
    S_{\alpha\beta} ={}& \frac{1}{eT} \frac{\mathcal{L}_{\alpha\beta}^1}{\mathcal{L}_{\alpha\beta}^0}, \\
    \kappa_{\alpha\beta} = {}& \frac{1}{e^2T}
    \left [ \frac{(\mathcal{L}_{\alpha\beta}^1)^2}{\mathcal{L}_{\alpha\beta}^0}
    - \mathcal{L}_{\alpha\beta}^2 \right ] .
\end{align}

\section{Computational Framework}

\subsection{Brillouin-zone interpolation and integration}

\begin{figure*}[t]
\includegraphics[width=1\textwidth]{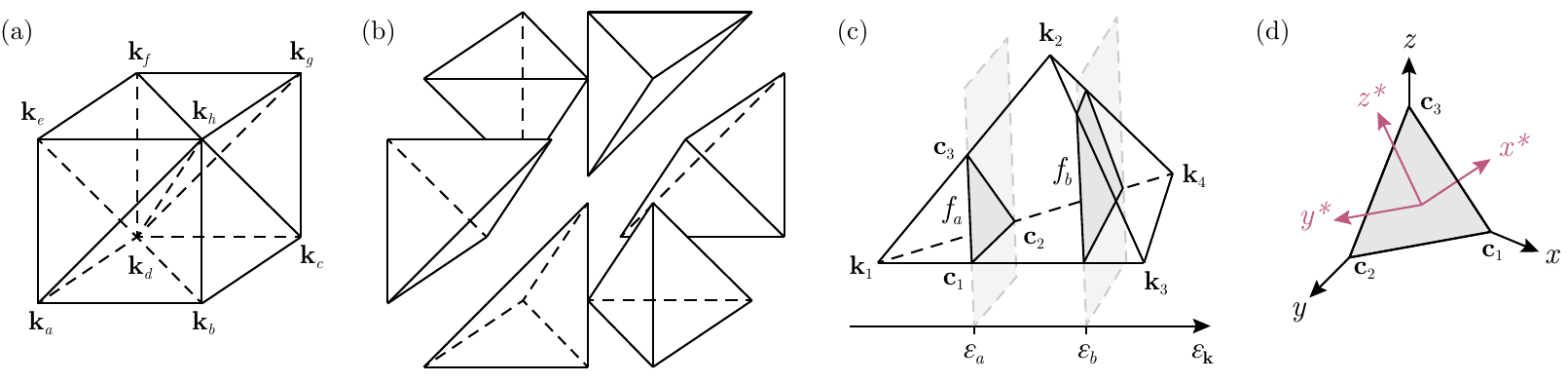}
\caption{\label{fig:tetrahedron} Schematic of the linear-tetrahedron method. (a) A $2\times2\times2$ $\mathbf{k}$-point submesh can be broken up into (b) six tetrahedra. Adapted from Ref.~\cite{blochl1994improved}. (c) The constant energy surfaces (light gray planes) defined by $\varepsilon_a$ and $\varepsilon_b$ intersect the tetrahedron to produce the cross sections $f_a$ (dark gray triangle) and $f_b$ (dark gray quadrangle). The triangular cross section $f_a$ is defined by the points $\mathbf{c}_1$, $\mathbf{c}_2$, and $\mathbf{c}_3$. The $\mathbf{k}$-points at the tetrahedron vertices have been numbered according to increasing energy, i.e., $\varepsilon_{\mathbf{k}_1} < \varepsilon_{\mathbf{k}_2} < \varepsilon_{\mathbf{k}_3} < \varepsilon_{\mathbf{k}_4}$. (d) Coordinate transformation from initial basis (black arrows) to transformed basis (pink arrows) that maps the cross section onto a 2D plane. The $x^*$ coordinates of all points on the cross section are zero.}
\end{figure*}

As described in the main text, we employ a combined Fourier-linear interpolation scheme when calculating scattering and transport properties.
Electronic eigenvalues --- calculated using density functional theory (DFT) on a coarse $\mathbf{k}$-point mesh --- are Fourier interpolated onto a denser mesh.
Fourier interpolation is performed using the \textsc{boltztrap2} software \cite{madsenBoltzTraP2ProgramInterpolating2018,madsenBoltzTraPCodeCalculating2006} which enforces symmetry using star functions and employs the Shankland algorithm to ensure that both quasi-particle energies and their derivatives (group velocities) are exactly reproduced \cite{euwemaCrystallineInterpolationApplications1969,koellingInterpolationEigenvaluesResultant1986,shanklandFourierTransformationSmooth2009}.
This approach aims to minimise the roughness function proposed in Ref.~\cite{pickettSmoothFourierInterpolation1988}.

Scattering rates are calculated on the Fourier interpolated $\mathbf{k}$-point mesh.
When calculating the partial decay rate, scattering is limited to the constant energy surface defined by $\varepsilon = \varepsilon_{n\mathbf{k}}$ in the case of elastic processes [Eq.~(\ref{eq:elastic-partial-rate})] and $\varepsilon = \varepsilon_{n\mathbf{k}} \pm \hbar\omega_\mathbf{q}$ for inelastic processes [Eq.~(\ref{eq:inelastic-partial-rate})].
Note that, in our implementation of polar optical phonon scattering we rely on a single dispersionless phonon mode, whose energy $\hbar\omega_\mathrm{po}$ is independent of $\mathbf{q}$.
Due to finite $\mathbf{k}$-point sampling, it is common replace the delta function in Eqs.~(\ref{eq:inelastic-partial-rate}) and (\ref{eq:elastic-partial-rate}) by Gaussian or Lorentzian functions with finite broadening.
This procedure has the effect that the calculated lifetimes will depend on the chosen broadening parameter.

An alternative approach is to employ the linear tetrahedron method to analytically integrate the scattering rates across the constant energy surface  \cite{blochl1994improved,lehmann1972numerical}.
In this method, the Brillouin zone is divided into tetrahedra [Figs.~\ref{fig:tetrahedron}(a) and \ref{fig:tetrahedron}(b)].
For each electronic band, the eigenvalues are obtained for the $\mathbf{k}$-points at the corners of the tetrahedra.
The constant energy surface defined by $\varepsilon_{n\mathbf{k}}$ intersects a tetrahedron if $\varepsilon_\mathrm{tetra}^\mathrm{min} < \varepsilon_{n\mathbf{k}} < \varepsilon_\mathrm{tetra}^\mathrm{max}$, where $\varepsilon_\mathrm{tetra}^\mathrm{min}$ and $\varepsilon_\mathrm{tetra}^\mathrm{max}$ are the minimum and maximum energies of the tetrahedron's vertices [Fig.~\ref{fig:tetrahedron}(c)].
Computing the intersections of $\varepsilon_{n\mathbf{k}}$ with all tetrahedra gives rise to a set of tetrahedron cross-sections that define the constant energy surface.
In the traditional implementation of the tetrahedron method, the integration for each tetrahedron is performed analytically after linearly interpolating the eigenvalues and matrix elements inside the tetrahedron.
As we note in the main text, this approach is only valid for matrix elements that show a linear dependence on $\mathbf{q}$.
For ionized impurity scattering, where the matrix element has a $1/\abs{\mathbf{q}}^2$ dependence, this assumption does not hold and results in severe overestimation of the scattering rate.

To overcome this limitation, we employ a modified linear-tetrahedron approach.
The constant energy surface is determined in the same manner as the tetrahedron method.
However, instead of analytically integrating within each tetrahedra, the tetrahedron cross sections (comprising the constant energy surface) are numerically resampled with hundreds of extra points.
By only computing additional $\mathbf{k}$-points that exactly satisfy the delta term in Eqs.~(\ref{eq:inelastic-partial-rate}) and (\ref{eq:elastic-partial-rate}), this allows for ``effective'' $\mathbf{k}$-point mesh densities that would be almost impossible to achieve with uniform $\mathbf{k}$-point sampling.
The scattering matrix elements are computed on the denser submesh by linear interpolation of the electronic wave functions $\psi_{n\mathbf{k}}$ and group velocities $\mathbf{v}_{n\mathbf{k}}$.
We note that the scattering wave vector $\mathbf{q}$ is a geometric term that is known exactly for all points on the submesh.
A primary advantage of this approach is that while the matrix elements cannot be linearly interpolated with $\mathbf{q}$, the constituent parameters (electronic wave functions and group velocities) \textit{are} linearly interpolatable.

In order to resample the constant energy surface, the tetrahedron cross sections are projected onto a two-dimensional plane.
First, the $\mathbf{k}$-points that define the tetrahedron cross sections are identified.
These are the points at the intersection of the constant energy surface and tetrahedron boundary under the assumption that the band energies vary linearly between adjacent vertices in the tetrahedron [points labelled $\mathbf{c}$ in Fig.~\ref{fig:tetrahedron}(c)].
This results in three and four sets of $\mathbf{k}$-points for triangular and quadrilateral cross sections, respectively, termed $\mathbf{C}$.
The first basis vector for the new coordinate system, $\mathbf{B}$, is the vector normal to the plane of the cross section, namely
\begin{equation}
    \mathbf{b}_1 = \frac{\mathbf{c}_{2} - \mathbf{c}_{1}}{\abs{\mathbf{c}_{2} - \mathbf{c}_{1}}} \cross \frac{\mathbf{c}_{3} - \mathbf{c}_{1}}{\abs{\mathbf{c}_{3} - \mathbf{c}_{1}}}, \nonumber
\end{equation}
where $\mathbf{c}_1$ and $\mathbf{c}_2$ are the coordinates of the first and second vertices defining the cross section.
The second and third basis vectors are defined as
\begin{align}
    \mathbf{b}_2 ={}& \frac{\mathbf{c}_{2} - \mathbf{c}_{1}}{\abs{\mathbf{c}_{2} - \mathbf{c}_{1}}}, \nonumber \\ 
    \mathbf{b}_3 ={}& \mathbf{b}_2 \cross \mathbf{b}_1, \nonumber
\end{align}
The reciprocal space coordinates defining the cross section are transformed onto the new basis through
\begin{equation}
    \mathbf{c}_i^\mathrm{proj} = \mathbf{B}^{-1} \cdot \mathbf{c}_i. \nonumber
\end{equation}
In the new coordinate system, the first component of all coordinates will be the same, as all vertices lie on a plane.
The last two components of the coordinates define a two-dimensional (2D) projection of the cross section which can be resampled through numerical quadrature schemes [Fig.~\ref{fig:tetrahedron}(d)].
In the present work, we employ degree 50 Xiao--Gimbutas (containing 453 sample points, \cite{xiao2010numerical}) or Festa--Sommariva quadratures (454 points, \cite{festa2012computing}) for resampling triangular and quadrilateral tetrahedron cross-sections, respectively.
Resampling, including generating sample points and integration weights $w^\mathrm{res}_i$, is performed using the \textsc{quadpy} software \cite{nico_schlomer_2020_3786435}.
The set of sample points are transformed back into the original coordinate system through
\begin{equation}
    \mathbf{c}_i = \mathbf{B} \cdot \mathbf{c}^\mathrm{proj}_i. \nonumber
\end{equation}

\begin{figure*}[t]
\includegraphics[width=0.95\textwidth]{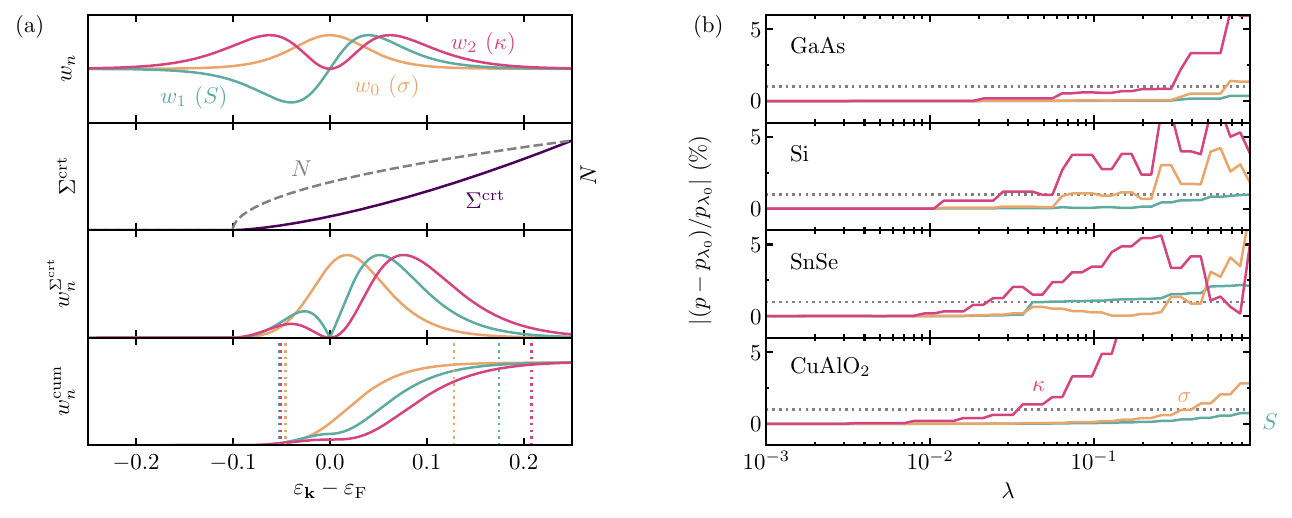}
\caption{\label{fig:fd-tol} (a) Procedure for obtaining the energy range in which to calculate scattering rates. The momentum coefficient weight $w_n$ for $n = 0, 1, 2$ (top panel) is scaled by the spectral conductivity $\Sigma^\mathrm{crt}$ to give $w_n^{\Sigma^\mathrm{crt}}$. The cumulative integral of the moment weights $w^\mathrm{cum}_n$ is used to determined the energy cutoffs (bottom panel). The dashed orange, teal, and pink lines give $\varepsilon_n^\mathrm{min}$ and $\varepsilon_n^\mathrm{max}$ for $n = 0, 1, 2$, respectively at $\lambda = 0.05$. The final values of $\varepsilon^\mathrm{min}$ and $\varepsilon^\mathrm{max}$ are taken as the smallest $\varepsilon_n^\mathrm{min}$ and largest $\varepsilon_n^\mathrm{max}$ values across all moments, respectively.
(b) Convergence of electronic transport properties $p$ as a function of $\lambda$ at \SI{300}{\kelvin} for GaAs, Si, SnSe, and \ce{CuAlO2}.
Absolute percentage difference from converged value $\abs{(p - p_{\lambda_0}) / p_{\lambda_0}}$ given for conductivity ($p = \sigma$, orange), Seebeck coefficient ($S$, teal), and electronic contribution to the thermal conductivity ($\kappa$, pink), respectively. $p_{\lambda_0}$ corresponds to the value of the transport properties at $\lambda = 0$ --- i.e., the scattering rates for all $\mathbf{k}$-points are calculated explicitly. Convergence within \SI{1}{\percent} is highlighted by a dashed gray line.}
\end{figure*}

The contribution of each tetrahedron to the constant energy surface is weighted by a geometric factor that accounts for the tetrahedron's shape in four dimensional space (reciprocal coordinates and energy space) \cite{lehmann1972numerical}.
Using the triple $\mathbf{r}_i$ contragradient to vertices of the tetrahedron $\mathbf{k}_i$
\begin{align}
    \mathbf{r}_i\mathbf{k}_i ={}& \delta_{ij}, \nonumber \\ 
    \mathbf{r}_1 ={}& \frac{\mathbf{k}_3 \cross \mathbf{k}_4}{\Omega}, \nonumber \\
    \mathbf{r}_2 ={}& \frac{\mathbf{k}_4 \cross \mathbf{k}_2}{\Omega}, \nonumber \\ 
    \mathbf{r}_3 ={}& \frac{\mathbf{k}_2 \cross \mathbf{k}_3}{\Omega}, \nonumber
\end{align}
where the $\mathbf{k}$-points have been numbered according to increasing energy, i.e., $\varepsilon_{\mathbf{k}_1} < \varepsilon_{\mathbf{k}_2} < \varepsilon_{\mathbf{k}_3} < \varepsilon_{\mathbf{k}_4}$, the tetrahedron weight is given by \cite{lehmann1972numerical}
\begin{equation}
    w^\mathrm{tet} = \abs{\sum_{i=2}^{4} \left (\varepsilon_{\mathbf{k}_i} - \varepsilon_{\mathbf{k}_1} \right ) \mathbf{r}_{i-1}}^{-1} . \nonumber
\end{equation}
We stress that this weight is distinct from the integration weights defined by \citet{blochl1994improved} in which the contragradient cancels when averaging over all adjacent tetrahedra. 
The final integration weights $w_i$ for the sample $\mathbf{k}$-point coordinates of each cross section are scaled by the tetrahedron weight to give $w_i = w^\mathrm{res}_i \cdot w^\mathrm{tet}$.

When evaluating the density of states
\begin{equation}
    N(\varepsilon) =  \sum_n \int \frac{\dd{\mathbf{k}}}{8\pi^3} 
    \delta{\left(\varepsilon - \varepsilon_{n\mathbf{k}} \right )},
\end{equation}
and the spectral conductivity in Eq.~(\ref{eq:spectral-cond}), we employ the traditional approach to the linear-tetrahedron method described by \citet{blochl1994improved}.
Specifically, we use the energy-dependent integration weights as described in Ref.~\cite{friedrich2019Tetrahedron} and elsewhere.
Unlike the partial decay rates $\tau_{n\mathbf{k}\rightarrow m\mathbf{k} + \mathbf{q}^{-1}}$, the final lifetimes $\tau_{n\mathbf{k}}$ vary smoothly across the Brillouin zone.
Accordingly, use of the linear-tetrahedron method can significantly improve the convergence of transport properties without issue.

\subsection{Optimization of scattering calculations}
\label{sec:lambda-cutoffs}

Under typically achievable carrier concentrations ($10^{16}$ to $10^{21}$\,cm$^2$/Vs) the Fermi level will sit close to either the conduction or valence band edge.
Accordingly, only $\mathbf{k}$-points that lie within a few hundred meV of the band edge will contribute to electronic transport.
It is therefore unnecessary to compute the electron lifetimes for all $\mathbf{k}$-points in the band structure, as most will have no impact on transport properties.
From the generalized transport coefficients $\mathcal{L}$ in Eq.~(\ref{eq:transport-moments}), it can be seen that each $\mathbf{k}$-point's contribution to the transport properties is scaled by a factor $(\varepsilon_{n\mathbf{k}} - \varepsilon_\mathrm{F})^n \left [ - \partial f^0_{n\mathrm{k}} / \partial \varepsilon_{n\mathbf{k}} \right ]$, which depends entirely on the energy of the state.
Accordingly, we have designed a procedure to assess which energy range is important for transport, illustrated in Fig.~\ref{fig:fd-tol}(a).
We begin by denoting the ``moment-coefficient weight'' as
\begin{equation}
    w_n(\varepsilon) = (\varepsilon_\mathrm{F} - \varepsilon)^n
    \left [ -\pdv{f^0}{\varepsilon} \right ],
\end{equation}
where the indices $n = 0, 1, 2$, correspond to the moments of $\mathcal{L}^n$ required to compute conductivity, Seebeck coefficient, and the electronic component of thermal conductivity, respectively.
This is weighted by the spectral conductivity $\Sigma^\mathrm{crt}$ under the assumption of a constant relaxation time [i.e., Eq.~(\ref{eq:spectral-cond}) with $\tau = 1$] to give
\begin{equation}
    w_n^{\Sigma^\mathrm{crt}}(\varepsilon) = \abs{w_n(\varepsilon)} \cdot \Sigma^\mathrm{crt}(\varepsilon).
\end{equation}
Finally, we compute the normalized cumulative integral of the weights according to
\begin{equation}
    w^\mathrm{cum}_n(\varepsilon) = \frac{\int_{-\infty}^\varepsilon w_n^{\Sigma^\mathrm{crt}}(\varepsilon^\prime) \dd{\varepsilon^\prime}}{
    \int w_n^{\Sigma^\mathrm{crt}}(\varepsilon^\prime) \dd{\varepsilon^\prime}}.
\end{equation}
We can then define a tuneable parameter $\lambda$ than controls the minimum and maximum energy ranges within which to calculate the scattering rates.
Namely,
\begin{align}
    \varepsilon^\mathrm{min}_n ={}& \argmin_\varepsilon \abs{w^\mathrm{cum}_n(\varepsilon) - \frac{\lambda}{2}}, \\
    \varepsilon^\mathrm{max}_n ={}& \argmin_\varepsilon \abs{w^\mathrm{cum}_n(\varepsilon) - \left [ 1 - \frac{\lambda}{2} \right ]},
\end{align}
where $\lambda$ can vary between $0$ (in which case $\varepsilon_n^\mathrm{min}$ and $\varepsilon_n^\mathrm{max}$ will be the minimum and maximum energies in the band structure) and $1$ (where $\varepsilon_n^\mathrm{min}$ and $\varepsilon_n^\mathrm{max}$ will be the same value).
A value of $\lambda = 0.1$, indicates that \SI{90}{\percent} of the integrated $w_n^{\Sigma^\mathrm{crt}}$ will be included in the energy range.
Alternatively put, a value of $\lambda = 0.1$ results in $\varepsilon_n^\mathrm{min}$ and $\varepsilon^\mathrm{max}$ taking the energies where $w_n^{\mathrm{cum}} = 0.05$ and 0.95, respectively.
The final energy range is given by $\varepsilon^{\min} = \min(\{\varepsilon^\mathrm{min}_n: n = 0, 1, 2\})$ and $\varepsilon^{\max} = \max(\{\varepsilon^\mathrm{max}_n: n = 0, 1, 2\})$.
The scattering rate is only calculated for states where $\varepsilon^\mathrm{min} \leq \varepsilon_{n\mathbf{k}} \leq \varepsilon^\mathrm{max}$, with the scattering rates of the remaining states set to the average value of the rates that have been calculated explicitly.
By setting $\lambda$ to an appropriate value, the scattering rates for $\mathbf{k}$-points outside the energy range will not impact the transport properties.

To demonstrate the impact of $\lambda$ and determine reasonable values to use in our calculations, we have investigated the convergence of the transport properties for GaAs, Si, SnSe, and \ce{CuAlO2} at \SI{300}{\kelvin} [Fig.~\ref{fig:fd-tol}(b)].
The conductivity, Seebeck coefficient, and electronic contribution to the thermal conductivity of all materials are converged to within than \SI{1}{\percent} by $\lambda = 0.02$.
In most cases, the Seebeck coefficient converges the fastest, most likely due to its weaker dependence on the scattering rate.
The electronic contribution to the thermal conductivity is the slowest property to converge, as expected from its reliance on a broader momentum coefficient weight.
If only the conductivity or Seebeck coefficient are of interest, a much larger value of $\lambda$ can be used. For example, using a $\lambda$ of 0.1 converges these properties to within \SI{1}{\percent}.
In our calculations, we employ a $\lambda$ of 0.05 which offers a reasonable trade-off between speed and convergence.
This property is controlled in our software implementation through the \verb|fd_tol| parameter.

\subsection{Software implementation}

\begin{figure}[t]
\includegraphics[width=\linewidth]{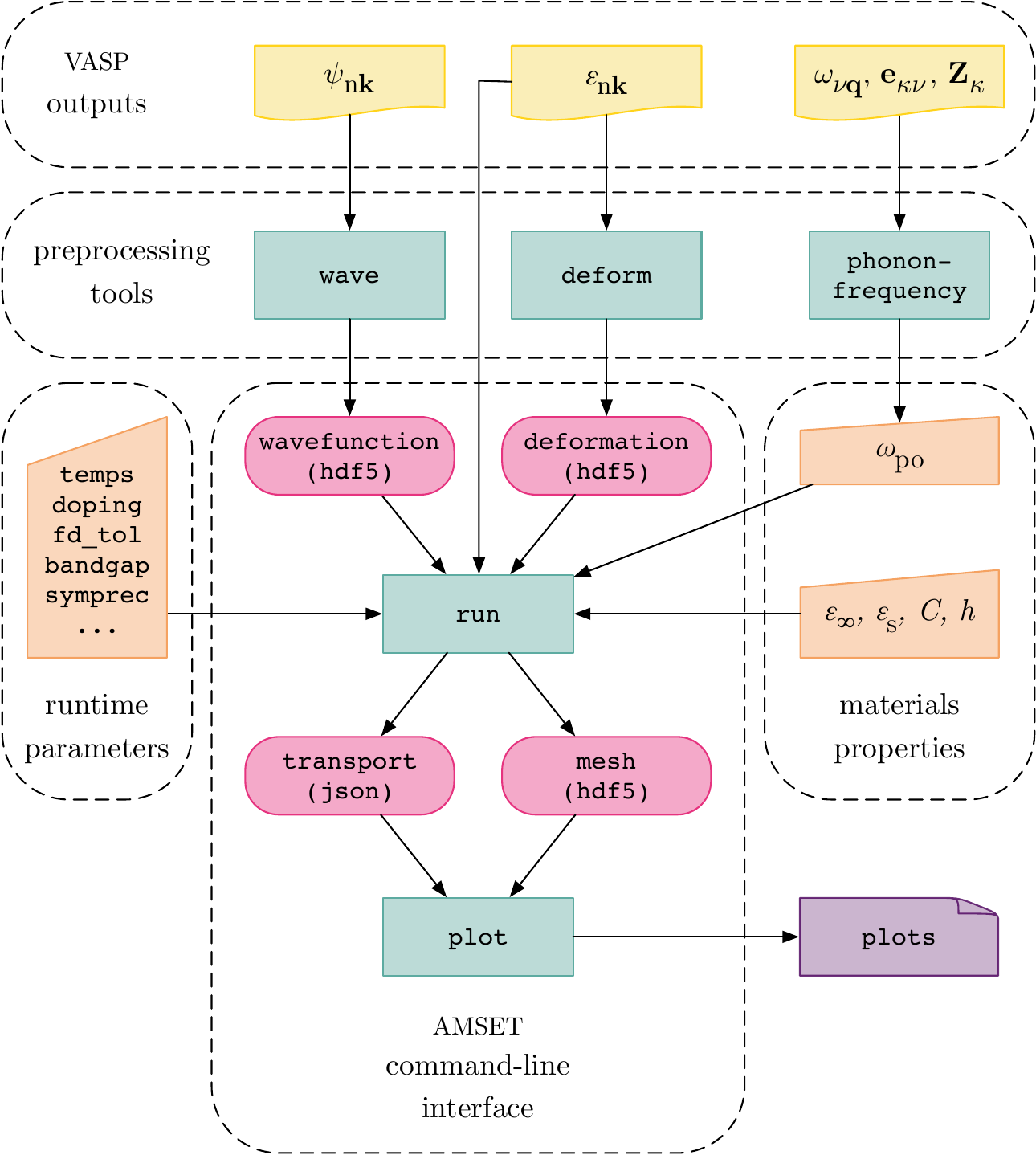}
\caption{\label{fig:amset} Schematic of the \textsc{amset} program indicating the typical inputs and outputs, command-line tools, and program flow.}
\end{figure}

An open-source implementation of the formalism, used to perform all calculations in this work, is released as a package called \textsc{amset} \cite{ganose2020amsetgithub}.
\textsc{amset} is freely available under a modified Berkeley Software Distribution (BSD) license.
The current version is developed and maintained using Git and is accessible at \url{https://hackingmaterials.lbl.gov/amset}. 
The code can be run on both high-performance computing clusters or personal computers.
\textsc{amset} is implemented in Python 3 and relies on several open-source libraries including pymatgen \cite{ongPythonMaterialsGenomics2013} for parsing \textsc{vasp} calculation outputs, \textsc{BoltzTraP2} \cite{madsen2006boltztrap,madsen2018boltztrap2} for Fourier interpolation of electronic eigenvalues and group velocities, \textsc{spglib} \cite{togo2018TextttSpglibSoftwarea} for symmetry analysis, \textsc{quadpy} \cite{nico_schlomer_2020_3786435} for numerical integration, and \textsc{matplotlib} \cite{hunter2007Matplotlib2D} for plotting.
The \textsc{NumPy} \cite{vanderwalt2011NumPyArray} and \textsc{SciPy} \cite{scipy1.0contributors2020SciPyFundamental} libraries are used extensively to minimize the cost of expensive matrix operations.
All-electron wave function coefficients are generated from the pseudo-wave functions using the \verb|MomentumMatrix| functionality of the \textsc{pawpyseed} package \cite{bystrom2019PawpyseedPerturbationextrapolation}.

\textsc{amset} can be used through either the  the command-line or a Python application programming interface (API).
A typical workflow, showing computational inputs and outputs, is illustrated in Fig.~(\ref{fig:amset}).
The primary inputs are \verb|vasprun.xml| and \verb|WAVECAR| \textsc{vasp} output files, calculated on a uniform $\mathbf{k}$-point mesh.
Additional settings, such as the materials parameters used to calculate scattering, the doping concentrations and temperatures to consider, and accuracy settings such as \verb|fd_tol|, can be specified in a separate file or as command-line arguments.
Information on all the available settings is provided on the \textsc{amset} website.
After obtaining the first principles inputs, two pre-processing steps are required.
Firstly, the all-electron wave function coefficients must be extracted from the \textsc{vasp} \verb|WAVECAR| file using the \verb|wave| tool.
Secondly, the ``effective-phonon-frequency'' should be calculated from phonon frequencies and eigenvectors, and the Born effective charges using the \verb|phonon-frequency| tool.
This process is described in more detail in Section \ref{sec:comp-methods}.
Scattering rates and transport properties are computed using the \verb|run| command.
The primary output is the \verb|transport| file, which by default contains the calculated mobility, Seebeck coefficient, and electronic contribution to the thermal conductivity in the JavaScript Object Notation (JSON) format.
The scattering rates, and interpolated eigenvalues and group velocities can be written to the \verb|mesh| file with the Hierarchical Data Format version 5 (HDF5) format \cite{folk2011overview} using the \verb|write_mesh| option.
Finally, the \verb|plot| command can be used to plot transport properties, lifetimes, and electron linewidths from the \verb|transport| and \verb|mesh| files.
The sumo package is used for plotting band structures \cite{ganose2018sumo}.

\subsection{Timing analysis}
\label{sec:timing}

\begin{figure}[t]
\includegraphics[width=\linewidth]{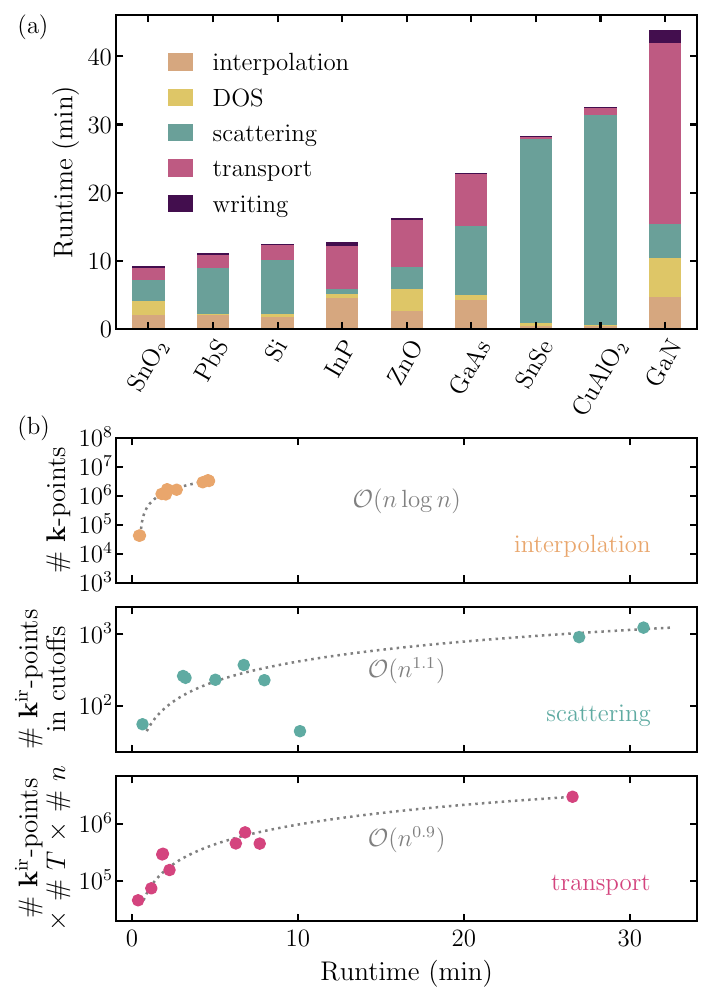}
\caption{\label{fig:timing} Timing analysis for running \textsc{amset} on a selection of materials in the test set. Calculations performed using the materials parameters in Table~\ref{tab:materials-parameters} and at the carrier concentrations and temperatures specified in Table~\ref{tab:mob-systems}. (a) The total runtime for each system, broken up into the different functions of the code. (b) Correlation between time and number (denoted by \#) of  $\mathbf{k}$-points for the interpolation, scattering, and transport routines.
$\mathbf{k}^\mathrm{ir}$ indicates the $\mathbf{k}$-points within the irreducible Brillouin zone. The number of temperatures and carrier concentrations are denoted by \# $T$ and \# $n$, respectively.
The computational complexity, provided in big O notation relative to the x-axis, is given in grey text and highlighted by dashed grey lines.
}
\end{figure}

A primary goal of the present approach is to be amenable to high-throughput computational workflows.
To investigate the computational requirements of the \textsc{amset} package, we have illustrated the time taken to calculate the scattering rates of several of the test materials in Fig.~\ref{fig:timing}(a).
All calculations were performed on a MacBook Pro with a quad core 2.9 GHz Intel Core i7 processor. 
The maximum time taken was \SI{42}{\min} for GaN, with most of the remaining materials completed in under \SI{20}{\min}.
To understand which portions of the code are the most computationally demanding, we have broken down the results into the time taken to:~(i) perform Fourier interpolation of electronic eigenvalues, (ii) compute the density of states through the tetrahedron method, (iii) obtain the scattering rates, (iv) calculate transport properties, and (v) write the output data to disk.
We note, the benchmarks were performed with the \verb|write_mesh| option enabled, so the output includes the scattering rates and interpolated band structure.
In general, writing the output data takes the least amount of time relative to the other functions of the code.
The breakdown for the rest of the computational steps depends strongly on the material and run time parameters, with most of the time spent calculating the scattering rates or transport properties.

To understand the scaling performance of \textsc{amset} with interpolation density, we have investigated the correlation of runtime with number of $\mathbf{k}$-points.
We find there is not a simple correlation between the total number of $\mathbf{k}$-points and total runtime.
Instead, each function of the code shows different scaling behaviour.
The interpolation routines show $\mathcal{O}(n \log n)$ scaling (where $n$ is the total number of $\mathbf{k}$-points in the dense mesh), which is consistent with the time complexity of the fast Fourier transform algorithm.
The time taken to compute scattering does not correlate well with total number of $\mathbf{k}$-points.
This is primarily as we only compute the scattering rates for the $\mathbf{k}$-points which fall within the energy cutoffs defined by the $\lambda$ parameter (see Section~\ref{sec:lambda-cutoffs}).
In addition, we use the symmetry of the reciprocal lattice to limit our calculations to the $\mathbf{k}$-points in the irreducible Brillouin zone (denoted $\mathbf{k}^\mathrm{ir}$-points).
The timing of the scattering routines correlates with the number of irreducible $\mathbf{k}$-points that fall within the energy cutoffs, exhibiting a $\mathcal{O}(n^{1.3})$ scaling complexity.
We note that, while the scattering rate is only calculated for the irreducible $\mathbf{k}$-points within the energy cutoffs, the scattering rate for each state requires integrating the partial decay rates over the full Brillouin zone and not just the irreducible part.
The time taken to compute transport properties correlates to the number the number of irreducible k-points multiplied by the number of carrier concentrations and temperatures included in the calculation, with a $\mathcal{O}(n^{0.9})$ scaling complexity.
The primary expense when computing transport properties is generating the energy-dependent tetrahedron integration weights used to obtain the spectral conductivity.

The total time to obtain transport properties is dominated by the calculation of the first-principles inputs (materials parameters and band structure calculation). In Table \ref{tab:timing-dft}, we provide the full timing information (in core hours) required to calculate all materials parameters used in this work.
In Fig.~(1) in the main text, we compare these times against DFPT+Wannier calculations performed using \textsc{quantum espresso} and \textsc{epw}.
In Table \ref{tab:timing-epw} we provide the full breakdown of the DFPT+Wannier calculations, including the references from which the timing information and mobility was extracted.

\begin{table}
\caption{Time required to obtain first-principles inputs given in core hours. Calculations were performed as described in the Computational Methodology. We note that the DFPT calculation listed here is performed only for a single $\mathbf{q}$-point at $\Gamma$ and is used to obtain the effective phonon frequency, static and high-frequency dielectric constants, and piezoelectric constants rather than the matrix elements $g(\mathbf{k}, \mathbf{q})$.
Static+NSCF (non self-consistent field) refers to a single point calculation on the relatively dense DFT $\mathbf{k}$-point meshes listed in Table \ref{tab:band-gaps}. Deformation and elastic refer to the total time required to calculate the deformation potential and elastic tensors}
\label{tab:timing-dft}
\begin{ruledtabular}
\begin{tabular}{lrrrrr}
Material & Static+NSCF &  Deformation &  DFPT &  Elastic & Total \\
\colrule
         GaAs &    0.30 &         0.91 &   9.33 &    10.31 &   20.85 \\
          GaN &    1.75 &        15.76 &  13.82 &    32.68 &   64.01 \\
          InP &    0.55 &         4.91 &   9.01 &     4.48 &   18.94 \\
          ZnS &    1.46 &        13.11 &   7.11 &    10.21 &   31.88 \\
         ZnSe &    1.47 &        13.19 &   7.90 &    10.24 &   32.79 \\
          CdS &    1.16 &        10.40 &  14.88 &    18.80 &   45.24 \\
         CdSe &    1.12 &        10.07 &  16.61 &    17.28 &   45.08 \\
         CdTe &    0.93 &         8.33 &   8.11 &     5.84 &   23.22 \\
          GaP &    1.48 &        13.36 &   8.43 &    10.24 &   33.52 \\
  MAPbI$_{3}$\footnotemark[1] &    6.67 &        40.03 & 901.12 &    65.03 & 1012.85 \\
          SiC &    2.57 &        23.16 &   3.23 &    17.69 &   46.66 \\
          PbS &    0.53 &         4.80 &   7.31 &     3.67 &   16.31 \\
    SnO$_{2}$ &    1.34 &        12.03 &  14.11 &    18.08 &   45.56 \\
          ZnO &    1.77 &        15.92 &  11.79 &    31.64 &   61.12 \\
         SnSe &    1.90 &        17.14 &  48.48 &    24.00 &   91.53 \\
  CuAlO$_{2}$ &    1.88 &        16.93 &  25.60 &    34.07 &   78.49 \\
           Si &    2.65 &         7.96 &   2.62 &     8.78 &   22.01 \\
 Ba$_{2}$BiAu &    1.78 &        16.05 &   9.16 &     5.55 &   32.49 \\
       NbFeSb &    1.21 &        12.15 &   5.52 &     6.15 &   25.04 \\
\end{tabular}
\end{ruledtabular}
\footnotetext{MA = \ce{CH3NH3}}
\end{table}

\begin{table}
\caption{Time required to obtain electron mobility using DFPT+Wannier, as implemented in \textsc{quantum espresso} (DFPT to obtain $g(\mathbf{k}, \mathbf{q})$ portion) and \textsc{epw} (Wannier interpolation and scattering portion) in core hours. References are given to the publications in which the timing information and mobility results are reported}
\label{tab:timing-epw}
\begin{ruledtabular}
\begin{tabular}{lrrrr}
Material & DFPT &  Scattering & Total & Refs. \\
\colrule
 Ba$_{2}$BiAu &    7000 & 2500 & 9500 & \cite{park2019HighThermoelectric} \\
       NbFeSb &    4600 & 2600 & 7200 & \cite{zhou2018LargeThermoelectric,samsonidze2018accelerated} \\
\end{tabular}
\end{ruledtabular}
\end{table}

\subsection{Reproducing the Brooks--Herring model of impurity scattering}

A primary advantage of the present approach is that it allows, for the first time, evaluation of ionized impurity scattering in anisotropic multi-band systems. Most modern computational evaluations of impurity scattering instead employ the closed-form Brooks--Herring formula \cite{brooks1951scattering,herring1956transport}. We will not reproduce the full derivation here but refer the reader to the excellent introduction provided in Ref.~\cite{chattopadhyay1981ElectronScatteringa}.
In this approach, the scattering matrix element
\begin{equation}
g_{nm}(\mathbf{k}, \mathbf{q}) =
     \frac{n_\mathrm{ii}^{1/2} Z e }{\epsilon_\mathrm{s}}
    \frac{1}{\left | \mathbf{q} \right | ^2 + \beta^2},
\label{eqn:element_ii}
\end{equation}
where $n_\mathrm{ii}$ and $Z$ are the concentration and charge of the charge of the impurities, $\epsilon_\mathrm{s}$ is the static dielectric constant, and $\beta$ is the inverse screening length given by Eq.~(\ref{eq:inv-screening}), is analytically integrated for a single parabolic band \cite{brooks1951scattering,herring1956transport}.
Under the assumption of complete overlap between the states the $n\mathbf{k}$ and $m\mathbf{k}+\mathbf{q}$, the resulting energy-dependent lifetime can be written
\begin{equation}
 \tau^{-1}_\mathrm{BH}(\varepsilon) = \frac{n_\mathrm{ii} Z^2 e^4 G(b)}{
     \pi 16 \sqrt{2}  \sqrt{m^*_\mathrm{d}} \epsilon_\mathrm{s}^2
 } \varepsilon^{-3/2},
\label{eq:bh-lifetime}
\end{equation}
where $m^*_\mathrm{d}$ is the density of states effective mass, $\epsilon_0$ is the vacuum permittivity, $G(b) = \ln(b+1) - b/(b+1)$, and $b=8m^*_\mathrm{d} \varepsilon / \hbar^2 \beta^2$.
Further integration of the energy-dependent lifetime yields the well-known Brooks--Herring mobility formula
\begin{equation}
 \mu_\mathrm{BH} = \frac{128 \sqrt{2 \pi} \epsilon_\mathrm{s}^2 (k_\mathrm{B} T)^{3/2}}{e^3  Z^2 \sqrt{m^*_\mathrm{d}} n_\mathrm{ii}  G(b)}.
\end{equation}

\begin{figure}[t]
\includegraphics[width=0.95\linewidth]{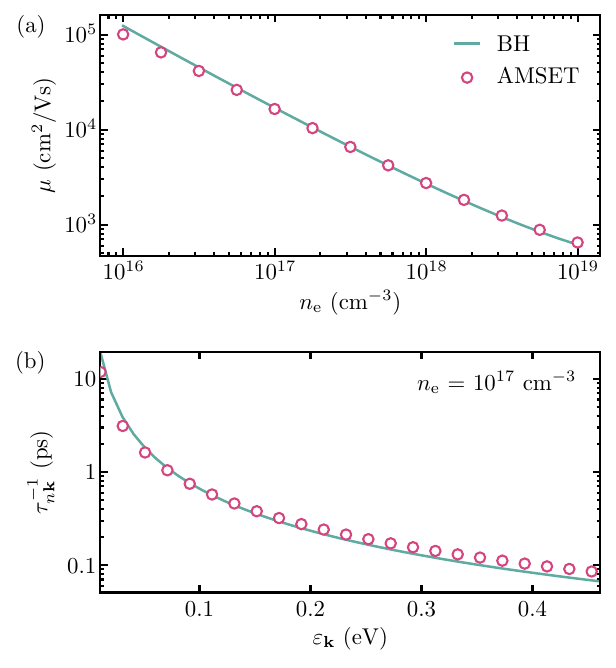}
\caption{\label{fig:bh}Comparison of the (a) mobility and (b) carrier lifetime between AMSET and the analytical Brooks--Herring formulas for a parabolic band structure. Results calculated at a temperature of \SI{500}{\kelvin}.
}
\end{figure}

To validate our implementation of ionized impurity scattering, we have generated a model parabolic electronic structure according to 
\begin{align}
\varepsilon_\mathbf{k} ={}& \frac{\hbar^2 \abs{\mathbf{k}}^2}{2m^*_\mathrm{d}}, \\
\mathbf{v}_\mathbf{k} ={}& \frac{\hbar \abs{\mathbf{k}}}{m^*_\mathrm{d}},
\end{align}
where $\varepsilon_\mathbf{k}$ and $\mathbf{v}_\mathbf{k}$ are the energy and group velocity at wave vector $\mathbf{k}$, respectively.
We calculated the ionized impurity scattering rate and resulting mobility using the AMSET package and Brooks--Herring formulas, parameterized according to $Z = 1$, $m^*_\mathrm{d} = 0.2$\,$m_0$, $\epsilon_\mathrm{s} = 20$\,$\epsilon_0$, $n_\mathrm{ii}$ = \SIrange{1E16}{1E19}{\per\centi\meter\cubed}, and $T$ = \SI{500}{\kelvin}.
A comparison between the two approaches is presented in Fig.~(\ref{fig:bh}).
Close agreement is observed for the both the mobility and carrier lifetime, indicating our approach is accurately reproducing the Brooks--Herring results.

\begin{figure}[t]
\includegraphics[width=0.95\linewidth]{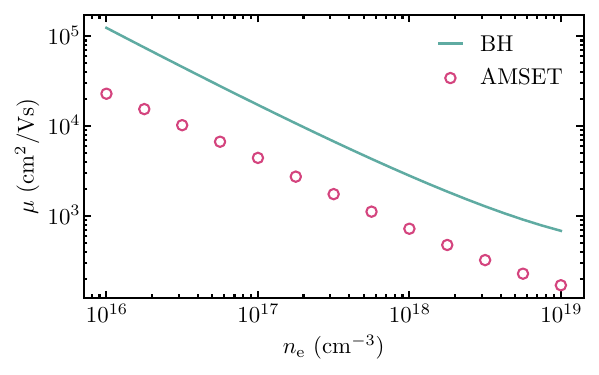}
\caption{\label{fig:bh-silicon}Comparison of the electron mobility between AMSET and the analytical Brooks--Herring formulas for a parameterized Silicon-like band structure. Results calculated at a temperature of \SI{500}{\kelvin}.}
\end{figure}

The Brooks--Herring formula is known to lead to inaccurate results for non-parabolic band structures or systems with multiple valleys.
To demonstrate this, we compare our method against Brooks--Herring on an idealized Silicon band structure, as parameterized in Refs.~\cite{dresselhaus1955CyclotronResonance} and \cite{peter2010fundamentals} and using the experimental effective masses according to
\begin{equation}
\begin{aligned}
\varepsilon_\mathbf{k} =&{} \frac{\hbar^2 (k_x - k_{0,x})^2}{2 m^*_\parallel} + \frac{\hbar^2 (k_y - k_{0,y})^2}{2 m^*_\perp} \\ &{} + \frac{\hbar^2 (k_z - k_{0,z})^2}{2 m^*_\perp},
\end{aligned}
\end{equation}
where $m^*_\parallel = 0.98$\,$m_0$, $m^*_\perp = 0.19$\,$m_0$, and $\mathbf{k}_0$ denotes the wave vectors of the conduction band minima.
The Brooks--Herring mobility is calculated using the harmonic mean of the effective masses, namely $3 / (m^{-1, *}_\parallel + 2m^{-1,*}_\perp ) = 0.26$\,$m_0$.
As can be seen in Fig.~(\ref{fig:bh-silicon}), the Brooks--Herring mobility is considerably over estimated by almost an order of magnitude relative to the mobility computed by AMSET.
This agrees well with empirical investigations into the mobility of Silicon that have noted the overestimation of the Brooks--Herring result \cite{norton1973ImpurityLattice}.

\section{First-principles inputs}

\subsection{Computational methodology}
\label{sec:comp-methods}

\begin{table*}[t]
\caption{Materials parameters used to compute scatterings rates. $\mathbf{C}$ is the elastic tensor in Voigt notation, with the unit GPa. $\boldsymbol{\epsilon}_\mathrm{s}$ and $\boldsymbol{\epsilon}_\mathrm{\infty}$ are the static and high-frequency dielectric constants in $\epsilon_0$. $\mathbf{D}^\mathrm{vb}$ and $\mathbf{D}^\mathrm{cb}$ are the absolute deformation potentials at the valence and conduction band edge, respectively. $d$ is the dimensionless piezoelectric coefficient. $\omega_\mathrm{po}$ is the effective polar phonon frequency given in THz. For all tensor properties, components that are not explicitly listed are zero}
\label{tab:materials-parameters}
\begin{ruledtabular}
\begin{tabular}{lrrrrrrrrrrrrrrrrrrrrrrrr}
Material &  $C_{11}$ & $C_{22}$ & $C_{33}$ & $C_{44}$ & $C_{55}$ & $C_{66}$ & $C_{12}$ & $C_{13}$ & $C_{23}$ & $\epsilon_{\mathrm{s},11}$ & $\epsilon_{\mathrm{s},22}$  & $\epsilon_{\mathrm{s},33}$ & $\epsilon_{\mathrm{\infty},11}$ & $\epsilon_{\mathrm{\infty},22}$ & $\epsilon_{\mathrm{\infty},33}$ & $D^\mathrm{vb}_{11}$ & $D^\mathrm{vb}_{22}$ & $D^\mathrm{vb}_{33}$  & $D^\mathrm{cb}_{11}$ & $D^\mathrm{cb}_{22}$ & $D^\mathrm{cb}_{33}$ & $d$ & $\omega_\mathrm{po}$ \\
\colrule
GaAs & 99 & 99 & 99 & 51 & 51 & 51 & 41 & 41 & 41 & 10.3 & 10.3 & 10.3 & 12.2 & 12.2 & 12.2 & 1.2 & 1.2 & 1.2 & 8.6 & 8.6 & 8.6 & 0.0 & 8.2 \\
GaN & 325 & 325 & 358 & 90 & 90 & 107 & 112 & 78 & 78 & 5.9 & 5.9 & 6.1 & 10.5 & 10.5 & 11.7 & 8.2 & 8.2 & 8.5 & 15.0 & 15.0 & 15.2 & 0.0 & 19.0 \\
InP & 87 & 87 & 87 & 42 & 42 & 42 & 46 & 46 & 46 & 13.2 & 13.2 & 13.2 & 16.5 & 16.5 & 16.5 & 1.6 & 1.6 & 1.6 & 5.7 & 5.7 & 5.7 & 0.0 & 10.3 \\
ZnS & 96 & 96 & 96 & 46 & 46 & 46 & 55 & 55 & 55 & 5.9 & 5.9 & 5.9 & 9.4 & 9.4 & 9.4 & 0.5 & 0.5 & 0.5 & 7.0 & 7.0 & 7.0 & 0.1 & 7.9 \\
ZnSe & 82 & 82 & 82 & 37 & 37 & 37 & 47 & 47 & 47 & 7.3 & 7.3 & 7.3 & 10.7 & 10.7 & 10.7 & 0.8 & 0.8 & 0.8 & 5.7 & 5.7 & 5.7 & 0.0 & 5.9 \\
CdS & 80 & 80 & 85 & 15 & 15 & 17 & 45 & 37 & 37 & 6.0 & 6.0 & 6.1 & 9.8 & 9.8 & 10.4 & 0.3 & 0.3 & 0.8 & 2.4 & 2.4 & 3.2 & 0.2 & 6.5 \\
CdSe & 66 & 66 & 72 & 13 & 13 & 15 & 36 & 31 & 31 & 8.8 & 8.8 & 8.7 & 12.4 & 12.4 & 13.0 & 0.2 & 0.2 & 0.6 & 2.4 & 2.4 & 2.9 & 0.1 & 4.7 \\
CdTe & 47 & 47 & 47 & 19 & 19 & 19 & 30 & 30 & 30 & 9.3 & 9.3 & 9.3 & 12.8 & 12.8 & 12.8 & 0.1 & 0.1 & 0.1 & 2.9 & 2.9 & 2.9 & 0.0 & 3.9 \\
GaP & 125 & 125 & 125 & 65 & 65 & 65 & 52 & 52 & 52 & 10.6 & 10.6 & 10.6 & 12.9 & 12.9 & 12.9 & 0.4 & 0.4 & 0.4 & 13.0 & 13.0 & 13.0 & 0.0 & 10.3 \\
MAPbI$_3$\footnotemark[1] & 43 & 55 & 53 & 10 & 20 & 11 & 33 & 31 & 17 & 31.7 & 97.5 & 66.7 & 5.1 & 5.2 & 5.1 & 4.0 & 4.7 & 4.0 & 2.8 & 3.1 & 2.5 & 0.0 & 2.6 \\
SiC & 382 & 382 & 382 & 241 & 241 & 241 & 126 & 126 & 126 & 7.0 & 7.0 & 7.0 & 10.3 & 10.3 & 10.3 & 5.6 & 5.6 & 5.6 & 3.4 & 3.4 & 3.4 & 0.2 & 23.2 \\
PbS & 121 & 121 & 121 & 20 & 20 & 20 & 18 & 18 & 18 & 15.6 & 15.6 & 15.6 & 277.7 & 277.7 & 277.7 & 1.6 & 1.6 & 1.6 & 1.0 & 1.0 & 1.0 & 0.0 & 6.6 \\
SnO$_2$ & 376 & 215 & 215 & 178 & 84 & 84 & 127 & 127 & 136 & 4.9 & 4.6 & 4.6 & 10.5 & 13.6 & 13.6 & 0.9 & 1.5 & 1.5 & 10.7 & 13.2 & 13.2 & 0.0 & 10.4 \\
ZnO & 188 & 188 & 205 & 37 & 37 & 39 & 109 & 92 & 92 & 3.8 & 3.8 & 3.8 & 10.5 & 10.5 & 11.4 & 7.6 & 7.6 & 8.2 & 9.1 & 9.1 & 9.5 & 0.0 & 11.2 \\
SnSe & 30 & 39 & 67 & 12 & 28 & 14 & 13 & 28 & 8 & 16.9 & 15.3 & 18.7 & 32.3 & 27.1 & 46.3 & 13.8 & 15.9 & 14.7 & 11.2 & 9.8 & 14.5 & 0.0 & 3.2 \\
CuAlO$_2$ & 294 & 294 & 509 & 39 & 39 & 102 & 90 & 103 & 103 & 6.1 & 6.1 & 4.6 & 10.8 & 10.8 & 7.3 & 2.5 & 2.5 & 4.8 & 7.1 & 7.1 & 10.2 & 0.0 & 14.0 \\
Si & 144 & 144 & 144 & 75 & 75 & 75 & 53 & 53 & 53 & 13.0 & 13.0 & 13.0 & 13.0 & 13.0 & 13.0 & 6.5 & 1.1 & 1.1 & 8.1 & 0.5 & 0.5 & 0.0 & 0.0 \\
Ba$_2$BiAu & 69 & 69 & 69 & 17 & 17 & 17 & 18 & 18 & 18 & 37.2 & 37.2 & 37.2 & 22.2 & 22.2 & 22.2 & 3.0 & 3.0 & 3.0 & 2.9 & 2.9 & 2.9 & 0.0 & 1.8 \\
NbFeSb & 309 & 309 & 309 & 67 & 67 & 67 & 95 & 95 & 95 & 44.7 & 44.7 & 44.7 & 24.7 & 24.7 & 24.7 & 1.1 & 1.1 & 1.1 & 0.6 & 1.6 & 1.6 & 0.0 & 7.4  \\
\end{tabular}
\end{ruledtabular}
\footnotetext{MA = \ce{CH3NH3}}
\end{table*}

First-principles calculations were performed using Kohn-Sham DFT \cite{hohenbergInhomogeneousElectronGas1964,kohnSelfConsistentEquationsIncluding1965} as implemented in the Vienna \textit{ab initio} Simulation Package (\textsc{vasp}) \cite{kresseInitioMoleculardynamicsSimulation1994,Kresse1996a,Kresse1996}.
All \textit{ab initio} inputs were computed within the generalized-gradient approximation (GGA) \cite{perdew1986AccurateSimple} using the Perdew-Burke-Ernzerhof (PBE) exchange--correlation functional \cite{perdew1996generalized}.
Calculations were performed in a plane-wave basis set with scalar relativistic psueodpoentials and with the interactions between core and valence electrons described using the projector augmented-wave method (PAW) \cite{blochlProjectorAugmentedwaveMethod1994,Kresse1999}.
The set-up, submission, and management of first-principles calculations was handled using the \textsc{atomate} workflow management software with the default parameters of version \verb|0.8.3| \cite{mathewAtomateHighlevelInterface2017,AtomateV02018}.
The plane-wave energy cutoff was set to \SI{520}{\electronvolt}.
Structure optimization was performed using the standard pymatgen \verb|MPRelaxSet| with a reciprocal $\mathbf{k}$-point density of \SI{64}{\textbf{k}\mbox{-}points\per\cubic\angstrom} \cite{ongPythonMaterialsGenomics2013}.
The uniform non-self-consistent calculations used as input to the scattering calculations were run with a reciprocal $\mathbf{k}$-point density of \SI{1000}{\textbf{k}\mbox{-}points\per\cubic\angstrom}.
Spin--orbit interactions were included for calculations on \ce{CH3NH3PbI3} as they were necessary to obtain the correct band ordering at the conduction band minimum.

Piezeoelectric constants, and static and high-frequency dielectric constants were computed using density functional perturbation theory (DFPT) based on the method developed and by \citet{baroni1986InitioCalculation} and adapted to the PAW formalism by \citet{Gajdos2006}.
Elastic constants were obtained through the stress-strain approach detailed in Ref.~\cite{dejongChartingCompleteElastic2015}.
These calculations were automated using the \verb|piezeoelectric_constant|, \verb|dielectric_constant|, and \verb|elastic_constant| preset workflows available in \textsc{atomate} \cite{mathewAtomateHighlevelInterface2017}.

Absolute volume deformation potentials were calculated in the manner proposed by \citet{wei1999PredictedBandgap}.
The deformation potential describes the change in energy of the bands with change in volume and was calculated as $\mathbf{D}_{n\mathbf{k}} = \delta \varepsilon_{n\mathbf{k}} / \delta S_{\alpha\beta}$ where $\mathbf{S}$ is the uniform stress tensor.
We average the deformation potential over contraction (\SI{-0.5}{\percent}) and expansion (+\SI{0.5}{\percent}) of the lattice.
Furthermore, we calculate the full deformation potential tensor by computing the deformation for each component of the strain tensor.
To account for shifts in the average electrostatic potential between deformed cells, we align the eigenvalues to the energy level of the deepest core state \cite{wei1999PredictedBandgap}.
We note that, in practice, even the reference energy levels can shift upon strain, leading to a small degree of error in the deformation potentials for non-covalent crystals \cite{resta1990AbsoluteDeformation,li2009RevisedInitio}.

The ``effective phonon frequency'' used in the calculation of polar-optical phonon scattering was determined from the phonon frequencies $\omega_{\mathbf{q}\nu}$ (where $\nu$ is a phonon branch and $\mathbf{q}$ is a phonon wave vector) and eigenvectors $\mathbf{e}_{\kappa\nu}(\mathbf{q})$ (where $\kappa$ is an atom in the unit cell).
In order to capture scattering from the full phonon band structure in a single phonon frequency, each phonon mode is weighted by the dipole moment it produces according to
\begin{equation}
    w_{\nu} = \sum_\kappa \left [ \frac{1}{M_\kappa \omega_{\mathbf{q}\nu}} \right]^{1/2}
    \times \left[ \mathbf{q} \cdot \mathbf{Z}_\kappa^* \cdot \mathbf{e}_{\kappa\nu}(\mathbf{q}) \right ]
\end{equation}
where $\mathbf{Z}_\kappa^*$ is the Born effective charge.
This naturally suppresses the contributions from transverse-optical and acoustic modes in the same manner as the more general formalism for computing Fr\"olich based electron-phonon coupling \cite{verdi2015frohlich,sjakste2015wannier}.
The weight is calculated only for $\Gamma$-point phonon frequencies and averaged over the unit sphere scaled by 0.01 to capture both the polar divergence at $\mathbf{q} \rightarrow 0$ and any anisotropy in the dipole moments.
The effective phonon frequency is calculated as the weighted sum over all $\Gamma$-point phonon modes according to
\begin{equation}
    \omega_\mathrm{po} = \frac{\omega_{\Gamma\nu} w_{\nu}}{\sum_{\nu} w_\nu}.
\end{equation}
We have released an open source tool \verb|phonon-frequency| as part of the \textsc{amset} package that automates this computation from \textsc{vasp} calculation outputs.

\subsection{Materials parameters}

All materials parameters were computed from first-principles in the manner described in the Computational Methodology.
A summary of the materials parameters used to compute carrier scattering rates is provided in Table \ref{tab:materials-parameters}.
We have additionally employed the rigid scissor approximation such that band gaps match those calculated using the hybrid HSE06 exchange--correlation functional.
Table \ref{tab:band-gaps} gives the band gaps and $\mathbf{k}$-point meshes employed in our calculations.
Furthermore, we report the range of temperatures and carrier concentrations at which mobility and Seebeck coefficients are computed in Tables \ref{tab:mob-systems} and \ref{tab:seeb-systems}.

\begin{table}[t]
\caption{Band gaps and $\mathbf{k}$-point meshes used to compute scatterings rates.  $\varepsilon^\mathrm{HSE}_\mathrm{g}$ and $\varepsilon^\mathrm{exp}_\mathrm{g}$ are the band gaps calculated using the HSE06 functional and taken from experiment, respectively, with the references given in square brackets. The coarse $\mathbf{k}$-point mesh of the electronic band structures computed using density functional theory (DFT) are compared to the dense mesh obtained through Fourier interpolation.}
\label{tab:band-gaps}
\begin{ruledtabular}
\begin{tabular}{lllcc}
  \multicolumn{3}{l}{} &
  \multicolumn{2}{c}{$\mathbf{k}$-point mesh} \\ \cline{4-5}
Material &  $\varepsilon_\mathrm{g}^\mathrm{HSE}$ (eV) &  $\varepsilon_\mathrm{g}^\mathrm{exp}$ (eV) & DFT & Interpolated \\
\colrule
GaAs  & 1.33 \cite{kimEfficientBandStructure2010} & 1.52 \cite{vurgaftman2001BandParameters} & $17\times17\times17$ & $143\times143\times143$ \\
GaN   & 3.06 \cite{stroppa2009UnravelingJahnTeller} & 3.26 \cite{vurgaftman2001BandParameters} & $20\times20\times12$ & $183\times183\times97$  \\
InP  & 1.48 \cite{kim2009AccurateBand} & 1.42 \cite{vurgaftman2001BandParameters} & $16\times16\times16$ & $151\times151\times151$ \\
ZnS & 3.22\footnotemark[1] & 3.72 \cite{tran1997PhotoluminescenceProperties} & $18\times18\times18$ & $133\times133\times133$ \\
ZnSe & 2.24\footnotemark[1] & 2.82 \cite{mang1994two} & $17\times17\times17$ & $99\times99\times99$ \\
CdS  & 2.12\footnotemark[1] & 2.48 \cite{ninomiya1995OpticalProperties} & $15\times15\times9$& $87\times87\times47$ \\
CdSe & 1.46\footnotemark[1] & 1.73 \cite{ninomiya1995OpticalPropertiesa} & $15\times15\times9$ & $87\times87\times47$ \\
CdTe & 1.34\footnotemark[1] & 1.48 \cite{lemasson1982FreeExcitons} & $15\times15\times15$ & $89\times89\times89$ \\
GaP & 2.37\footnotemark[1] & 2.24 \cite{foster1965ELECTROLUMINESCENCEBAND} & $18\times18\times18$ & $105\times105\times105$ \\
MAPbI$_3$\footnotemark[2] & 2.43\footnotemark[1] & 1.63 \cite{doi:10.1021/acsphotonics.6b00139} & $7\times4\times6$ & $51\times33\times47$ \\
SiC & 2.35\footnotemark[1] & 2.36 \cite{liu2013OpticalMechanical} & $22\times22\times22$ & $125\times125\times125$ \\
PbS & 0.84 \cite{walsh2011EffectsReduced} & 0.37 \cite{madelung2012semiconductors} & $16\times16\times16$ & $119\times119\times119$ \\
\ce{SnO2} & 2.88 \cite{tran2017ImportanceKinetic} & 3.60 \cite{batzill2005SurfaceMaterials} & $19\times13\times13$ & $135\times91\times91$  \\
ZnO & 2.55 \cite{hinuma2014BandAlignment} & 3.37 \cite{madelung2012semiconductors} & $20\times20\times12$ & $145\times145\times77$ \\
SnSe & 1.10 \cite{huang2017FirstprinciplesStudy} & 0.90 \cite{soliman1995OpticalProperties} & $13\times13\times5$ & $51\times49\times17$ \\
\ce{CuAlO2} & 3.52 \cite{scanlon2010ConductivityLimits} & 2.97 \cite{tate2009OriginType} & $14\times14\times4$ & $57\times57\times13$ \\
Si & 1.15 \cite{hinuma2014BandAlignment} & 1.14 \cite{madelung2012semiconductors} & $18\times18\times18$ & $105\times105\times105$ \\
Ba$_2$BiAu & 0.88\footnotemark[1] & --- & $11\times11\times11$ & $41\times41\times41$ \\
NbFeSb & 1.26\footnotemark[1] & 0.51 \cite{he2016AchievingHigh} & $16\times16\times16$ & $45\times45\times45$ \\
\end{tabular}
\end{ruledtabular}
\footnotetext{This work.}
\footnotetext{MA = \ce{CH3NH3}}'
\end{table}

\begin{table}[t]
\sisetup{tight-spacing=true,range-phrase=--}
\caption{Summary of temperature and doping conditions used for computing electron mobility. References provided to Electron--Phonon Wannier (EPW) calculations and  experimental measurements performed at the same doping and temperature conditions, which are used in the comparison of electron mobilities in the main text and Supplemental Material}
\label{tab:mob-systems}
\begin{ruledtabular}
\begin{tabular}{lcrrrrc}
Material & Doping & $T$ (K) & $n$ (cm$^{-3}$) & Exp. & EPW  \\
\colrule
GaAs & $n$-type & 200--1000 & \num{3.0e13} & \cite{rode1971electron} & \cite{zhou2016InitioElectron} \\
GaAs & $p$-type &  300 & \numrange{3.0e13}{8.6e19} & \cite{jansak1972EffectMagnetic,hill1970ActivationEnergy} & --- \\
GaN & $n$-type & 150--500 &  \numrange{3.0e16}{5.5e16} & \cite{steigerwald1997IIINitride} & \cite{ponce2019HoleMobility} \\
InP & $n$-type & 150--700 & \num{1.5e16} & \cite{galavanov1970MechanismElectron} & --- \\
ZnS & $n$-type & 300--650 & \num{1.0e16} & \cite{kroger1956OpticalElectrical} & --- \\
ZnSe & $n$-type & 200--1300 & \numrange{4.0e14}{2.0e15}  & \cite{aven1971HighElectron,smith1969EvidenceNative} & --- \\
CdS & $n$-type & 100--400 & \num{5.0e15} & \cite{podor1971ElectronConcentration} & --- \\
CdSe & $n$-type & 150--1300 & \numrange{1.0e16}{1.e18} & \cite{btirmeister1967ElectricalProperties,smith1970HighTemperature} & --- \\
CdTe & $n$-type & 100--1200 & \numrange{5.4e14}{1.4e17} & \cite{segall1963ElectricalProperties,smith1970ElectricallyActive} & --- \\
CdTe & $p$-type & 550--1000 & \numrange{1.4e16}{6.7e16} & \cite{smith1970ElectricallyActive} & --- \\
GaP & $n$-type & 100--500 & \num{3.0e16} & \cite{taylor1968ElectricalOptical} & --- \\
MAPbI$_3$\footnotemark[1] & $n$-type & 100--350 & \num{1.0e14} & \cite{karakus2015PhononElectron,milot2015TemperatureDependent} & \cite{ponce2019OriginLow}\\
SiC & $n$-type & 100--850 & \numrange{3.7e15}{2.5e16} & \cite{shinohara1988GrowthHighMobility} & \cite{meng2019PhononlimitedCarrier} \\
PbS & $n$-type & 300--750 & \num{3.6e17} & \cite{petritz1955MobilityElectrons} \\
\ce{SnO2} & $n$-type & 300--700 & \num{1.0e17} & \cite{fonstad1971ElectricalProperties} & --- \\
ZnO & $n$-type & 300--1000 & \num{8.2e16} & \cite{hutson1957HallEffect}  & --- \\
SnSe & $p$-type & 300--600 & \num{3.0e17} & \cite{zhao2014UltralowThermal} & \cite{ma2018IntrinsicPhononlimited} \\
\ce{CuAlO2} & $p$-type &  300--430 & \numrange{1.3e17}{7.4e18} & \cite{tate2009OriginType} & --- \\
Si & $n$-type & 300 & \numrange{2.0e14}{4.4e18} & \cite{jacoboni1977ReviewCharge} & \cite{ponce2018towards}\\
Ba$_2$BiAu & $n$-type & 300 & \num{1e14} & \cite{park2019HighThermoelectric} & --- \\
NbFeSb & $p$-type & 300 & \num{2e20} & \cite{zhou2018LargeThermoelectric,samsonidze2018accelerated} & --- \\
\end{tabular}
\end{ruledtabular}
\footnotetext{MA = \ce{CH3NH3}}
\end{table}

\begin{table}[thpb]
\sisetup{tight-spacing=true,range-phrase=--}
\caption{Summary of temperature and doping conditions used for computing Seebeck coefficient. References provided to experimental measurements performed at the same doping and temperature conditions, which are used in the comparison of Seebeck coefficients in the main text and Supplemental Material}
\label{tab:seeb-systems}
\begin{ruledtabular}
\begin{tabular}{lcrrr}
Material & Doping & $T$ (K) & $n$ (cm$^{-3}$) & Exp.  \\
\colrule
GaAs & $n$-type & 400--750 & $3.5\times10^{17}$ & \cite{sutadhar1979ThermoelectricPower} \\
GaAs & $p$-type & 350--750 & $6.4\times10^{19}$ & \cite{amith1965ElectronPhonon} \\
GaN & $n$-type  & 100--300 & $1.3\times10^{19}$ & \cite{sulkowski2010TransportProperties} \\
InP & $n$-type  & 150--700 & $2.1\times10^{17}$ & \cite{kudman1964ThermalConductivity} \\
CdS & $n$-type & 130--300 & \num{2.8e15} & \cite{morikawa1965SeebeckEffect} \\
PbS & $n$-type  & 300--800 & $2.5\times10^{19}$ & \cite{wang2013HighThermoelectric} \\
\ce{SnO2} & $n$-type & 300--800 & $8.2\times10^{18}$ & \cite{morgan1966ElectricalProperties} \\
ZnO & $n$-type  & 200--1000 & $5.2\times10^{17}$ & \cite{tsubota1997ThermoelectricProperties} \\
SnSe & $p$-type & 300--600 & $3.0\times10^{17}$ & \cite{zhao2014UltralowThermal} \\
Si & $n$-type & 300 & \numrange{1e14}{1e19} & \cite{geballe1955SeebeckEffect,herring1954TheoryThermoelectric} \\
\end{tabular}
\end{ruledtabular}
\end{table}

\subsection{Experimental data}

In the main text, we calculate the mobility and Seebeck coefficient of 17 semiconductors and compare our results to experimental measurements.
Our set of test materials spans a range of chemistries and doping-polarities and contains both isotropic and anisotropic materials.
The set includes: (i) conventional semiconductors, Si, GaAs, GaN, GaP, InP, ZnS, ZnSe, CdS, CdSe, and SiC;~(ii) the thermoelectric candidate SnSe;~(iv) photovoltaic absorbers \ce{CH3NH3PbI3}, PbS, and CdTe;~and (iii) transparent conductors, \ce{SnO2}, ZnO, and \ce{CuAlO2}.
The reference samples are of the highest purity and crystallinity in order to minimize the mesoscopic effects of grain boundary scattering and crystallographic one-dimensional and two-dimensional defects (e.g., line dislocations, edge dislocations, and stacking faults).
We favor bulk crystals over thin films (which can exhibit surface effects that impact carrier transport, e.g., strain, oxidation, off-stoichiometries, and surface dipole moments), however, in some cases we use epitaxial single crystal films.
We also favor undoped or dilutely doped crystals (to less than \SI{0.5}{\percent} at.) to avoid the formation of secondary crystal phases and degenerate doping.
Lastly, we favor studies that look at a wide range of carrier concentrations and/or temperatures (greater than 300K).
In all cases, experimental mobility is measured via the DC Hall effect.
A summary of the reference data used in the comparisons against carrier mobility and Seebeck coefficient are provided in Tables \ref{tab:mob-systems} and \ref{tab:seeb-systems}.





\clearpage
\onecolumngrid
\section{Mobility results}

\subsection{Temperature and carrier dependent mobility}

\begin{figure}[H]
\includegraphics[width=\textwidth]{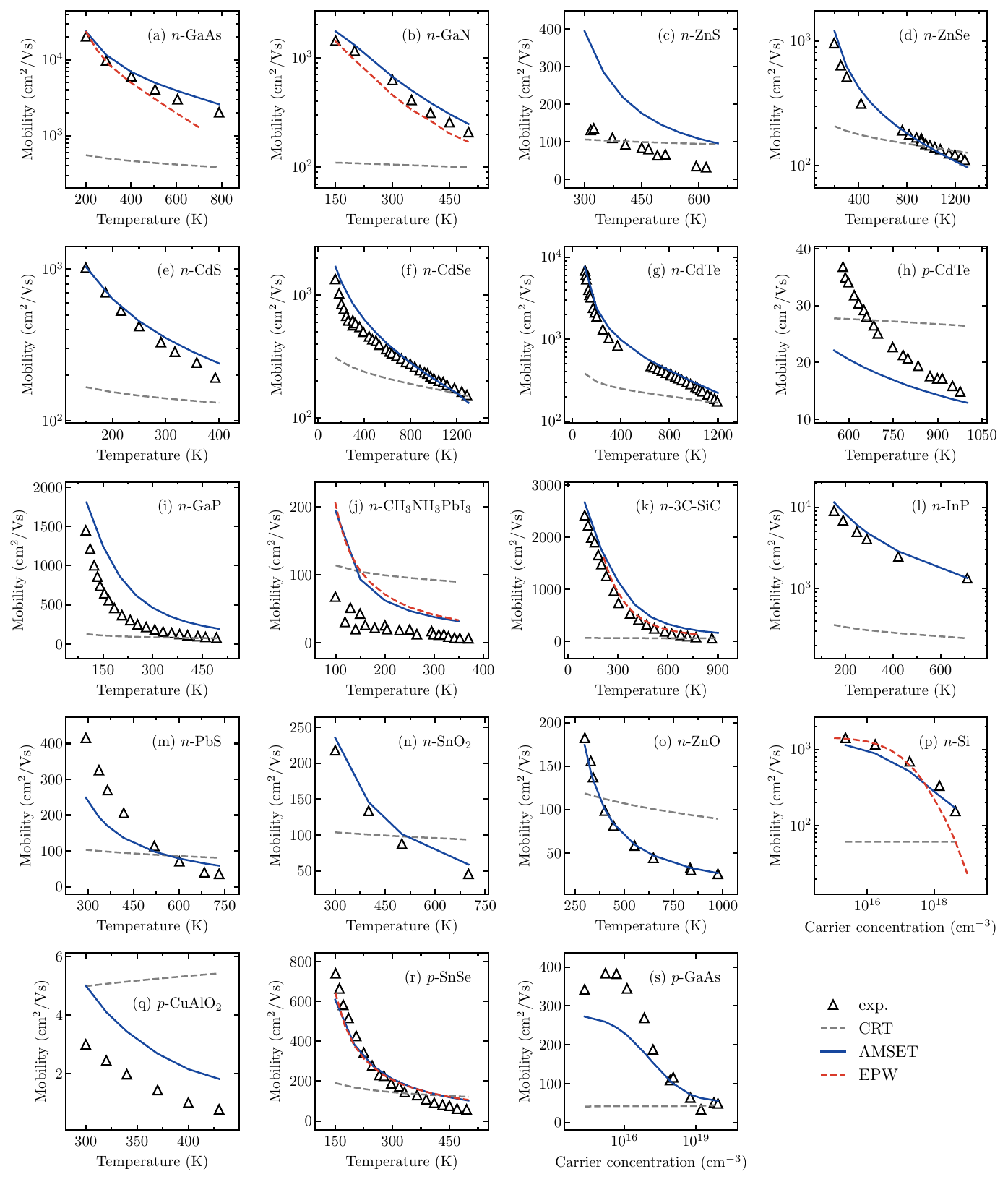}
\caption{\label{fig:mob-hse} Mobility against temperature or carrier-concentration for all test materials, computed using the HSE06 band gap.}
\end{figure}

\clearpage

\subsection{Scattering limited mobilities}

\begin{figure}[H]
\includegraphics[width=\textwidth]{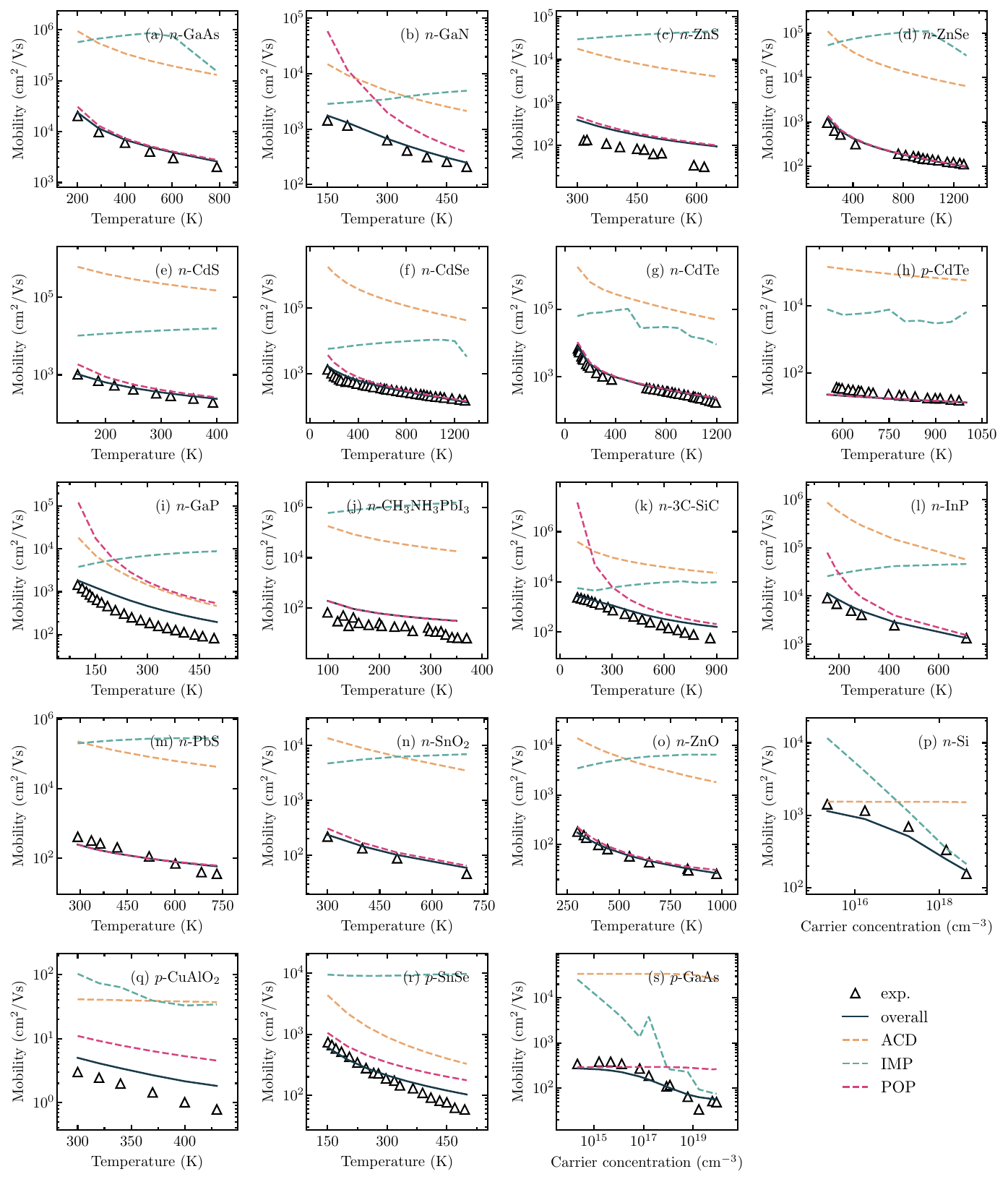}
\caption{\label{fig:mob-hse-scats} Mobility limited by different scattering mechanisms against temperature or carrier-concentration for all test materials, computed using the HSE06 band gap.}
\end{figure}
\clearpage

\subsection{Mobility calculated using the HSE06 functional}

\begin{figure}[H]
\includegraphics[width=0.85\textwidth]{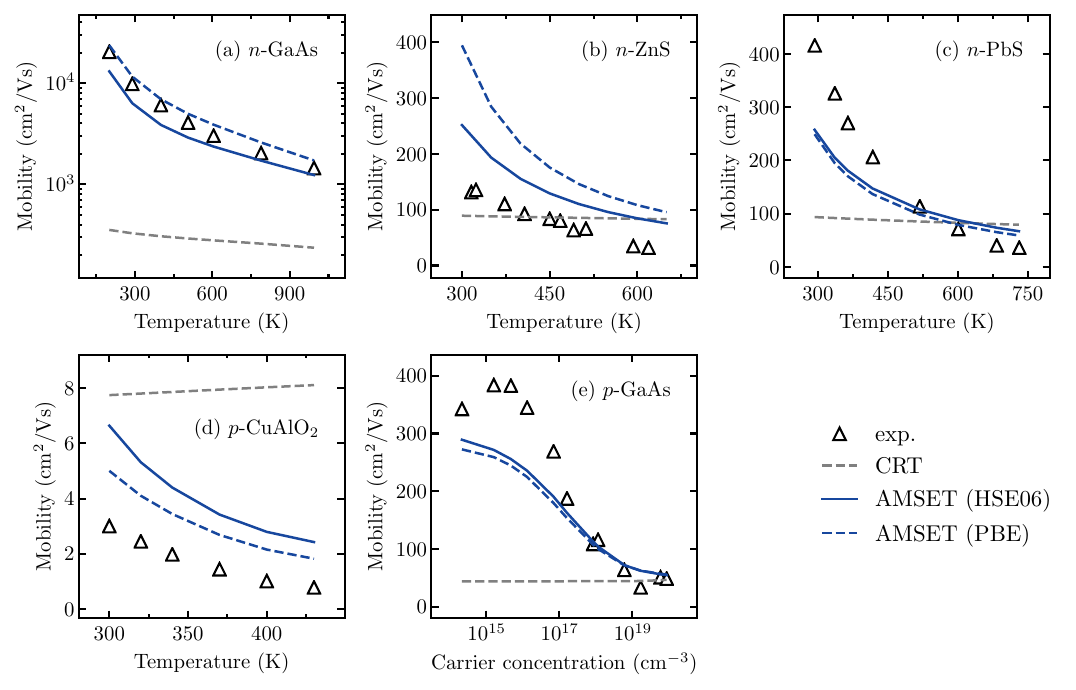}
\caption{\label{fig:mob-hse-vr-hse}Mobility against temperature or carrier-concentration for a set of test materials, computed using HSE06 electronic structures.}
\end{figure}
\clearpage

\section{Seebeck coefficient results}

\subsection{Temperature- and carrier concentration-dependent Seebeck coefficient}

\begin{figure}[H]
\includegraphics[width=\textwidth]{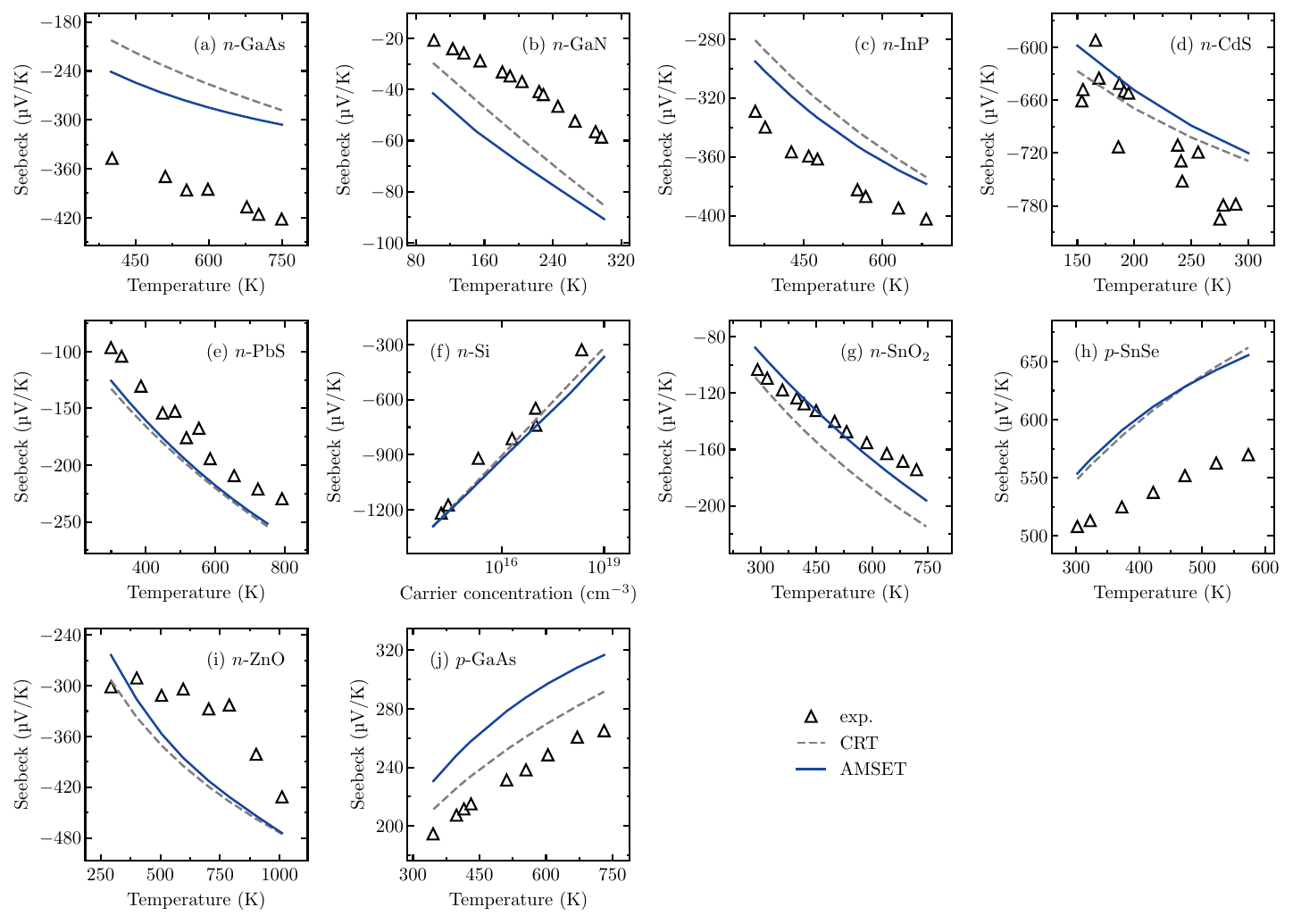}
\caption{\label{fig:seeb-hse} Seebeck coefficient against temperature for all test materials, computed using the HSE06 band gap.}
\end{figure}

\subsection{Seebeck coefficient calculated using the HSE06 functional}

\begin{figure}[H]
\includegraphics[width=\textwidth]{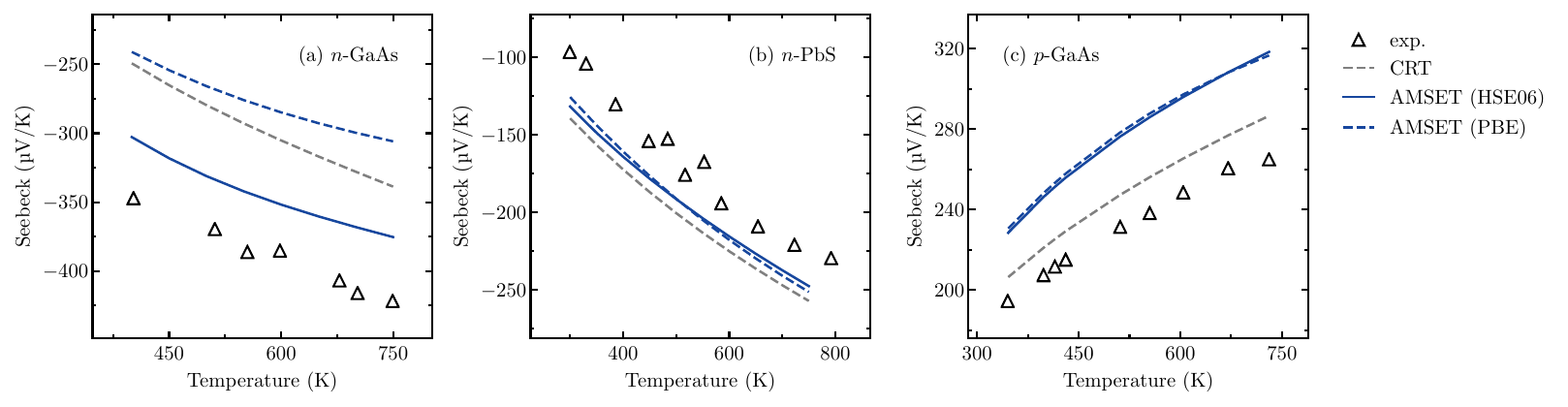}
\caption{\label{fig:seeb-hse-vr-hse}Seebeck coefficient against temperature for a set of test materials computed using HSE06 electronic structures.}
\end{figure}

\clearpage

\section{Scattering rate comparison}

\begin{figure}[H]
\centering
\includegraphics[width=0.8\textwidth]{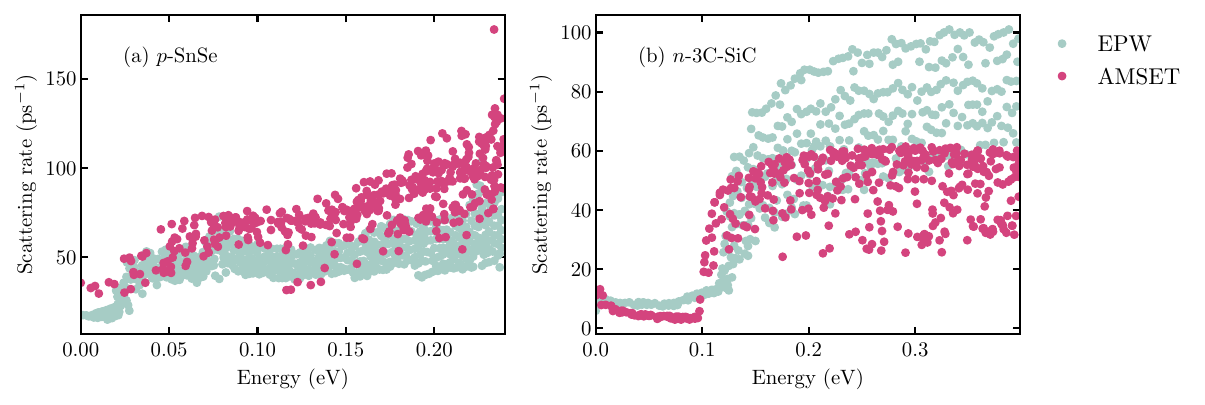}
\caption{\label{fig:hse-rates}Computed scattering rates compared against EPW calculations \cite{zhou2016InitioElectron,meng2019PhononlimitedCarrier,ponce2018towards,ma2018IntrinsicPhononlimited}. Results calculated at \SI{300}{\kelvin} using the the lowest carrier concentrations for each material given in Table. \ref{tab:mob-systems}}.
\end{figure}

\section{Comparison against CRT and EPW}

\begin{figure}[H]
\includegraphics[width=\textwidth]{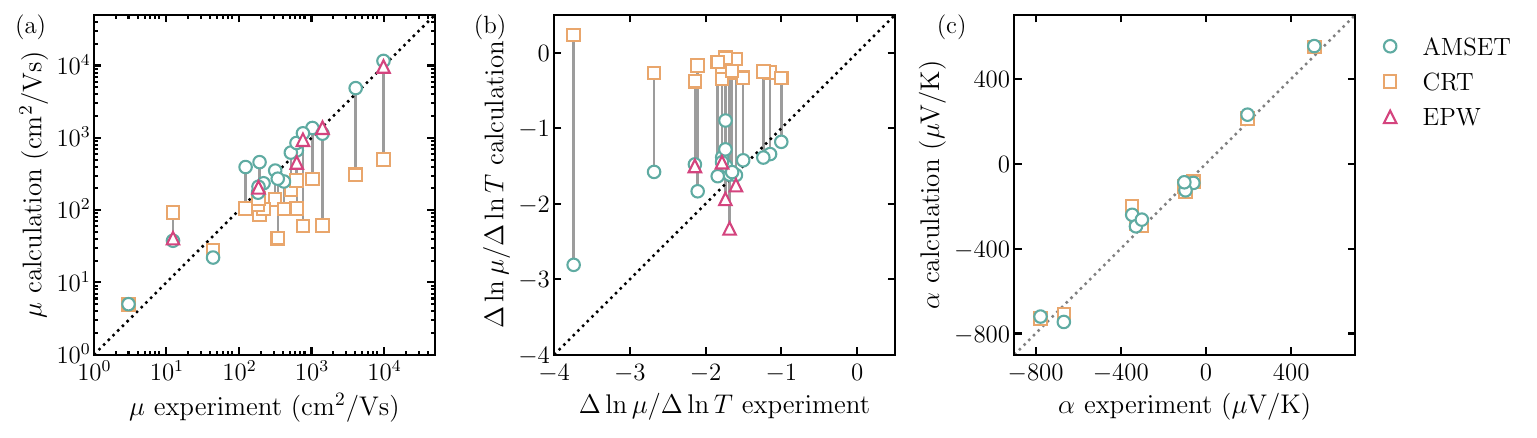}
\caption{\label{fig:summary-crta-epw} Comparison between AMSET, constant relaxation time approximation calculations, EPW calculations, and experiments for (a) carrier mobilities at \SI{300}{\kelvin} (b) the exponential temperature trend of carrier mobilities, and (c) Seebeck coefficients at \SI{300}{\kelvin}.}
\end{figure}
\clearpage

\section{Band structures}

\begin{figure}[H]
\includegraphics[width=\textwidth]{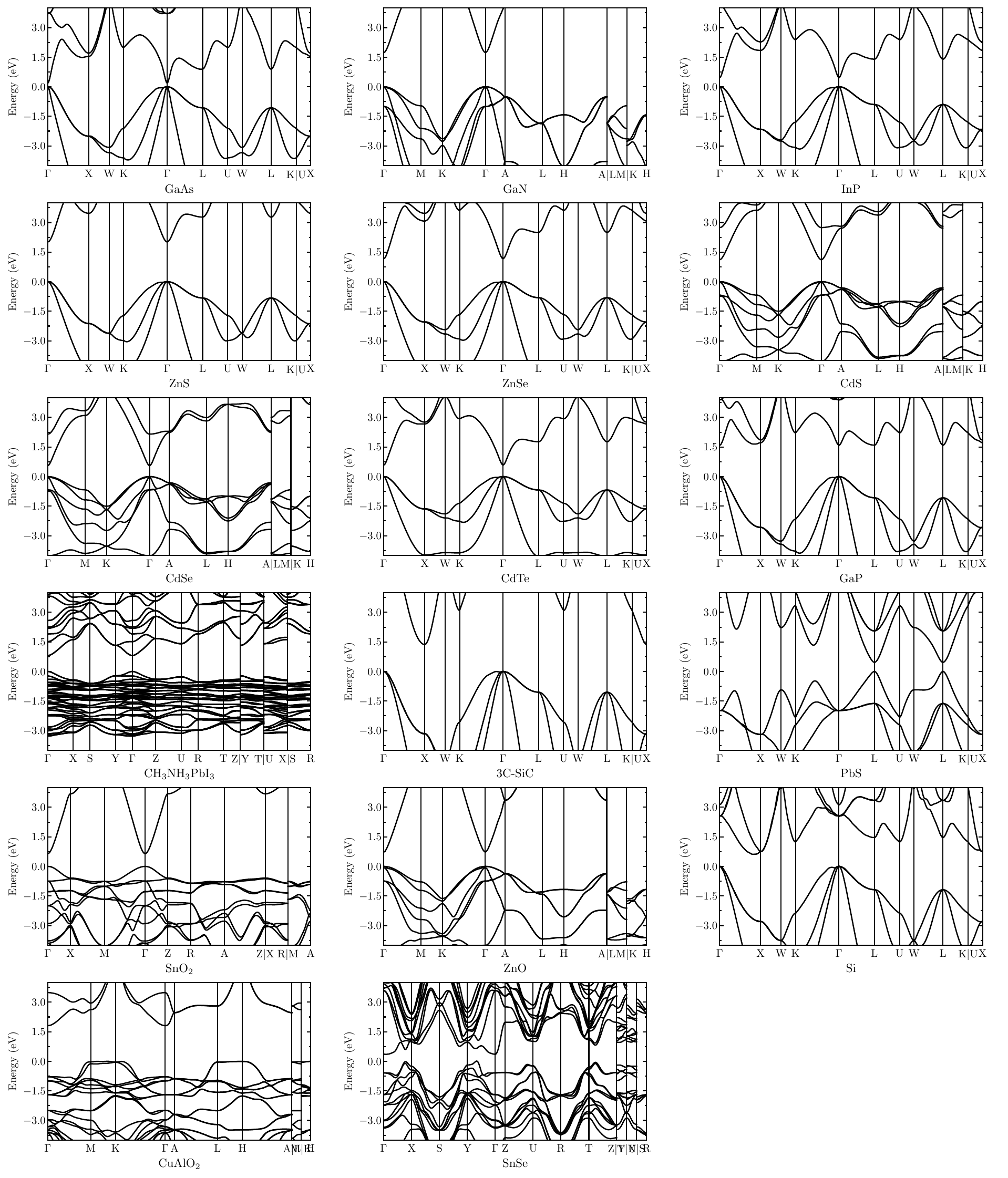}
\caption{\label{fig:band-structures}Band structures (pre-scissor operation) calculated using the PBE exchange--correlation functional, interpolated from a uniform $\mathbf{k}$-point mesh using the \textsc{boltztrap2} package.}
\end{figure}
\clearpage


\clearpage

\twocolumngrid
\bibliography{main}